\documentclass{article}

\usepackage[a4paper,margin=1in]{geometry}
\usepackage[authoryear]{natbib}

\DeclareFontShape{OT1}{cmr}{bx}{sc}{<-> ecxc1000}{}
\DeclareFontShape{OT1}{cmr}{m}{scit}{<->ssub*lmr/m/scsl}{}

\usepackage{amsmath}
\usepackage{amssymb}
\usepackage{amsthm}
\usepackage{collect}
\usepackage{graphicx}
\usepackage[frozencache=true,cachedir=minted-cache]{minted}
\usepackage{stmaryrd}
\usepackage{tikz-cd}

\usepackage[unicode,colorlinks=false,pdfborder={0 0 0},bookmarks=false]{hyperref}

\newcommand{\textalpha}{\texorpdfstring{\(\alpha\)}{α}}
\renewcommand{\textbeta}{\texorpdfstring{\(\beta\)}{β}}
\newcommand{\texteta}{\texorpdfstring{\(\eta\)}{η}}
\newcommand{\textlambda}{\texorpdfstring{\(\lambda\)}{λ}}
\newcommand{\textSigma}{\texorpdfstring{\(\Sigma\)}{Σ}}
\newcommand{\biarrow}{\raisebox{0.5pt}[0pt][0pt]{$\genfrac{}{}{0pt}{3}{\raisebox{-1pt}{$\rightarrow$}}{\raisebox{1pt}{$\leftarrow$}}$}}

\renewcommand{\P}{\textsc{p}}
\newcommand{\E}{\textsc{e}}
\newcommand{\coq}{R\textsc{ocq}}

\begin{document}

\title{Formal \P-Category Theory and Normalization by Evaluation in \coq{}}

\author{
    David G.\ Berry \\
    Dept of Computer Science and Technology, University of Cambridge \\
    (\href{mailto:David.Berry@cl.cam.ac.uk}{\texttt{David.Berry@cl.cam.ac.uk}})
    \and
    Marcelo P.\ Fiore \\
    Dept of Computer Science and Technology, University of Cambridge \\
    (\href{mailto:Marcelo.Fiore@cl.cam.ac.uk}{\texttt{Marcelo.Fiore@cl.cam.ac.uk}})
}

\date{\vspace{-\baselineskip}}

\maketitle

\begin{abstract}
  \noindent Traditional category theory is typically based on set-theoretic principles and ideas, which are often non-constructive.
  An alternative approach to formalizing category theory is to use \E-category theory, where hom sets become setoids.
  Our work reconsiders a third approach -- \P-category theory -- from \cite{cubric_normalization_1998} emphasizing a computational standpoint.
  We formalize in \coq{} a modest library of \P-category theory -- where homs become subsetoids -- and apply it to formalizing algorithms for normalization by evaluation which are purely categorical but, surprisingly, do not use neutral and normal terms.
  \cite{cubric_normalization_1998} establish only a soundness correctness property by categorical means; here, we extend their work by providing a categorical proof also for a strong completeness property.
  For this we formalize the full universal property of the free Cartesian-closed category, which is not known to have been performed before.
  We further formalize a novel universal property of unquotiented simply typed \textlambda-calculus syntax and apply this to a proof of correctness of a categorical normalization by evaluation algorithm.
  We pair the overall mathematical development with a formalization in the \coq{} proof assistant, following the principle that the formalization exists for practical computation.
  Indeed, it permits extraction of synthesized normalization programs that compute (long) \textbeta\texteta-normal forms of simply typed \textlambda-terms together with a derivation of \textbeta\texteta-conversion.
\end{abstract}

\theoremstyle{plain}
\newtheorem{theorem}{Theorem}
\newtheorem{lemma}[theorem]{Lemma}
\newtheorem{corollary}[theorem]{Corollary}

\theoremstyle{definition}
\newtheorem{definition}[theorem]{Definition}
\newtheorem{construction}[theorem]{Construction}
\newtheorem{example}[theorem]{Example}

\theoremstyle{remark}
\newtheorem*{remark}{Remark}

\numberwithin{theorem}{subsection}
\numberwithin{equation}{theorem}
\numberwithin{listing}{subsection}
\numberwithin{figure}{subsection}

\definecollection{rocq}

\section{Introduction}
\cite{cubric_normalization_1998} propose a categorical framework using \emph{partial equivalence relations}, which they termed \P-category theory, to overcome obstacles of quotienting in type theory; specifically, in the context of normalization for the simply typed \textlambda-calculus (STLC).

The word problem for STLC is a, if not \emph{the}, main problem in simple type theory.
It asks when two terms are \textbeta\texteta-convertible: \emph{i.e.}, when does the following hold?
\[ \Gamma \vdash t \cong_{\beta\eta} t' : \tau \]
The well-known, and standard, solution normalizes the terms so that they may be compared (decidably) using \textalpha-equivalence, reducing the problem to the following one.
\[ \Gamma \vdash \mathsf{nf}(t) \equiv_\alpha \mathsf{nf}(t') : \tau \]
Such a normalization function needs to produce a unique representative from the \textbeta\texteta-conversion class; \emph{i.e.}, a term \textbeta\texteta-convertible with the original term.
It must satisfy the following two properties, which we respectively call \emph{strong completeness}, and \emph{soundness}.
\[
    \frac{\Gamma \vdash t \cong_{\beta\eta} t' : \tau}{\Gamma \vdash \mathsf{nf}(t) \equiv_\alpha \mathsf{nf}(t') : \tau}
\quad \quad
    \frac{ }{\Gamma \vdash \mathsf{nf}(t) \cong_{\beta\eta} t : \tau}
\]
Note that strong completeness only demands a \emph{canonical} form for each \textbeta\texteta-conversion class that may not necessarily be \emph{normal} in any conventional sense.

The above creates a challenge for any semantical analysis of normalization functions: extensionally normalization is simply the identity; but, intensionally, it needs to compute normal forms.
Syntactic categorical models of the simply typed \textlambda-calculus all too readily quotient the set of terms by the equivalence relation of \textbeta\texteta-conversion, resulting in a normalization function being extensionally (a form of) the identity function and thereby being extensionally indistinguishable from the intensional identity function.

\cite{cubric_normalization_1998} note that their \P-category theory framework ``is constructive and thus permits extraction of programs from proofs'' and that it is ``fundamental to extracting a normalization algorithm''.
They further suggest that \P-category theory ``may be the appropriate way to develop category theory inside a constructive framework such as Martin-L\"of type theory''.
Finally, they stress: ``that the \P-category theory we use can be formalized (programmed) in Martin-L\"of type theory is one way of ensuring that our normalization function is indeed an algorithm''.
We realize, as well as extend, their vision by formalizing \P-category theory in \coq{} and reproduce their normalization algorithm in this computational setting.

\subsection{Contributions}
We present a complete formalization of the normalization function of \cite{cubric_normalization_1998} in \coq{}.
Moreover, we have formalized the full categorical freeness two-dimensional universal property for Cartesian closure (though not required for normalization) given in their work.
\cite{cubric_normalization_1998} prove that their normalization function is strongly complete by non-categorical means.
We present here a novel universal property of unquotiented syntax that allows for a new categorical proof of strong completeness by way of Artin-Wraith gluing.
Informally and intuitively this gluing construction sits between the gluing construction of \cite{fiore_2002,fiore_2022} and the work of \cite{cubric_normalization_1998}.
All our work has been formalized in \coq{} and we present concrete \coq{} definitions alongside abstract mathematical definitions.

\subsection{Related Work}
Our work covers both category theory and normalization by evaluation for the simply typed \textlambda-calculus, together with formalizations thereof.
Hence, we divide our comparison and contrastments with related work across these two aspects.

\subsubsection{Category Theory}
There have been many formalizations of category theory, for many proof assistants, each with their own peculiarities.
We summarize a few here for some brief comparisons and contrastments.

The formalization of category theory in \coq{} dates back to at least the work of \cite{huet_2000}, who use setoids for homs implementing \E-category theory.
They remark that subsetoids (for which they use the term `partial setoid') could be used instead, thereby implementing \P-category theory.
However, no further observations about this direction are made.

\cite{gross_2014}, report on their experience formalizing category theory in \coq{}.
They use a variety of techniques for their library; namely, dependently-typed morphisms, bundled records, and duality-centric design.
We follow them by using dependently-typed morphisms and bundled records.

More recently, \cite{hu_2021}, have implemented a category theory library for Agda.
In their library they use universe polymorphism, proof-relevant setoids, and duality-centric design.
We follow them by using the universe polymorphism available in \coq{} (which post-dates the work of \cite{gross_2014}).
Our work differs from theirs in two respects: by using proof-irrelevant subsetoids and \P-category theory.

The use of these techniques in \cite{hu_2021} allows for the double opposite of a category to be judgmentally equivalent with the original category rather than simply canonically categorically equivalent.
Our adoption of strict propositions in \coq{} provides this judgmental equivalence (and many others) for ``free'' and, as such, we do not extensively use duality-centric design.

Univalent category theory has also seen formalization efforts in both \coq{} and Agda such as the work of \cite{bauer_hott_2017}.
These formalizations, embracing the homotopical interpretation of type theory, allow for the assertion of axioms that are validated in their intended models when these principles are not provided by the respective proof assistant.
For example, function extensionality follows from the univalent setting and thus \cite{bauer_hott_2017} assume it as an axiom.
Our approach avoids the assumption of such axioms by explicitly tracking the uses of extensionality locally instead of assuming it as a global principle.

Category theory has also been formalized by \cite{mathlib_2020} in a non-constructive setting making use of quotients.
Their use of quotients rather than setoids contrasts heavily with our work.

\subsubsection{Normalization}
Normalization of STLC terms has two varieties: reduction-free and reductional.
The latter arises from the perspective of viewing STLC as a rewrite system, with a rewriting theory.
This work focuses on the former which has been studied most successfully using \emph{Normalization by Evaluation} (NbE).
We consider here only other work on simple type theories, rather than non-simple type theories containing polymorphism, type constructors, or dependent types.

Normalization by evaluation has its roots in the work of \cite{berger_nbe_1991}.
Their work was partially categorified by \cite{altenkirch_norm_1995}; although, that work still made use of \emph{ad-hoc} non-categorical techniques in order to overcome the intension \emph{versus} extension problem: ``the [...] normalization function only works on equivalence classes and is hence extensionally equal to the identity.
We introduce \emph{normal form objects} to overcome this problem''.
\cite{fiore_2002,fiore_2022} further categorified NbE to fully use Artin-Wraith gluing, rather than the \emph{ad-hoc} twisted gluing technique of \cite{altenkirch_norm_1995}.
These categorifications, ultimately, relied upon the syntactic notions of neutral and normal forms in order to state the algorithm and prove its correctness.
The use of \P-category theory in the work of \cite{cubric_normalization_1998} partially obviates the need for such syntactic considerations, since characterizing the resultant forms as normal inevitably requires the notions of normal and neutral forms.
They only prove some of the correctness properties given in \cite{fiore_2002,fiore_2022} categorically, and resort to non-categorical techniques for other proofs.

\cite{kovacs_2017} formalizes and proves correctness properties of NbE in Agda; although, minimally connecting to category theory.
In discussion on formalizing strong completeness, \cite{kovacs_2017} notes that partial equivalence relations could be used in the formalization but that this was found to be too inconvenient.
This contrasts with the work of \cite{cubric_normalization_1998} and our work, which centres on the use of partial equivalence relations.
In fact, we believe that the categorical nature of \cite{cubric_normalization_1998}, which we inherit, overcomes to a certain extent the inconvenience experienced by \cite{kovacs_2017}.
The structured framework of our categorical development guides precisely all that needs to be formalized, and provides an interface for abstract reasoning: once a categorical structure has been instantiated all general properties of the categorical structure are available at once, and instantiating the categorical structure is often much easier than proving the desired properties for specific concrete instances.

\subsection{Organization}
In section~\ref{Section:PCategoryTheory}, we present basic definitions in \P-category theory from both a pen-and-paper perspective and the perspective of the \coq{} formalization.
Moreover, in section~\ref{Section:PCategoricalStructure}, we present definitions of \P-category theory relevant for the \P-categorical analysis of the simply typed \textlambda-calculus.
In section~\ref{Section:SimplyTypedSyntax}, we describe our formalization of the simply typed \textlambda-calculus.
In section~\ref{Section:CategoricalStructure}, we connect the \P-categorical definitions with the simply typed \textlambda-calculus; and, in section~\ref{Section:ComputationalApplication}, we define the normalization function with two correctness properties.
Finally, in section~\ref{Section:Conclusions}, we summarize our work and results, and suggest avenues for further work.

\section{\P-Category Theory}
\label{Section:PCategoryTheory}

\P-Category theory is a form of category theory first proposed in the work of \cite{cubric_normalization_1998} to allow the separation of intensional and extensional behaviour for the extraction of a normalization algorithm.
Their choice of \P-category theory over \E-category theory was motivated by a separation of `data parts' from `property parts'.
Interestingly, this separation has some profound implications when comparing and contrasting completeness with cocompleteness.
In \E-category theory, completeness and cocompleteness in the \E-category of \E-sets are achieved by distinct strategies: the former by \textSigma-types and the latter by the equivalence relations associated to setoids.
However, in \P-category theory, completeness and cocompleteness in the \P-category of \P-sets are both achieved by the same strategy, namely by that of the partial equivalence relations associated to subsetoids.

Although \P-category theory can be given a formal categorical foundation by considering a form of enrichment, in this paper we follow a direct, and more practical and applied, presentation.

\subsection{\P-Sets}
We give some definitions for the PER (Partial Equivalence Relation) setting.
These basic constructions provide a flavour of the theory of \P-sets.

\begin{definition}[\texttt{PER}]
    A \emph{partial equivalence relation} (PER) is a symmetric and transitive relation.

    We use the addition of strict propositions by \cite{gilbert_2019} to \coq{} for the valuation of our PERs: this amounts to the \coq{} definition in listing~\ref{Listing:PER}.
    \begin{listing}
        \begin{minted}[tabsize=2,breaklines,frame=lines]{coq}
Record PER (A : Type) := {
  PER_rel : A -> A -> SProp;
  PER_symm : forall {x y}, PER_rel x y -> PER_rel y x;
  PER_trans : forall {x y z}, PER_rel x y -> PER_rel y z -> PER_rel x z;
}.
        \end{minted}
        \caption{PER \coq{} Definition}
        \label{Listing:PER}
    \end{listing}
\end{definition}

\begin{definition}[\texttt{PType}]
    In the \coq{} formalization we define a record \emph{\texttt{PType}} which bundles together a (carrier/underlying) type with a PER over the given type.
\end{definition}
By abuse of notation facilitated in \coq{} by a coercion, we will identify any given \texttt{PType} with its underlying carrier type.
Although, when we wish to emphasize that we are referring to the underlying carrier type (rather than the whole \texttt{PType} structure) we will use vertical bars around the name of the \texttt{PType}.
Furthermore, we will denote the associated PER by \texttt{\(\sim\)}, sometimes with a disambiguating subscript.

\begin{construction}[Discrete \texttt{PType}]
    Any \coq{} type, \(A\), has an associated discrete \texttt{PType}, \(A_=\), where:
    \begin{itemize}
        \item \(|A_=| \triangleq A\); and
        \item \(a \sim a' \triangleq a = a'\).
    \end{itemize}
\end{construction}

\begin{construction}[\texttt{PUnit}]
    The unit \texttt{PType} has a single, self-related element.
    This is implemented by using the unit type for the underlying type, and the terminal strict proposition for the associated PER.
\end{construction}

\begin{construction}[\texttt{PProd}]
    The product of two \texttt{PType}s, \(A\) and \(B\), is given by the following data:
    \begin{itemize}
        \item \(| A \times B | \triangleq |A| \times |B|\); and
        \item \((a, b) \sim_{A \times B} (a', b') \triangleq a \sim_A a' \wedge b \sim_B b'\).
    \end{itemize}
\end{construction}

\begin{construction}[\texttt{PArr}]
    The exponential of two \texttt{PType}s, \(A\) and \(B\), is given by the following data:
    \begin{itemize}
        \item \(| A \rightarrow B | \triangleq |A| \rightarrow |B|\); and
        \item \(f \sim_{A \rightarrow B} f' \triangleq \forall {a\, a'}. \, a \sim_A a' \Rightarrow f \, a \sim_B f' \, a'\).
    \end{itemize}
\end{construction}
Note that partiality lends itself to using the logical definition of when two functions are related.
Such a definition dates as far back as \cite{gandy_ext_1956}, see also \cite{hyland_2016}.

\begin{remark}
    The partiality afforded by omitting reflexivity offers some advantages over the setting with equivalence relations.
    To appreciate why this is the case consider that when using setoids/ERs, subtypes can only be formed by changing the underlying type to an appropriate \textSigma-type to restrict by the desired property.
    This mixes computational data (the base of the \textSigma-type) with a formalization property (the fibre of the \textSigma-type).
    This lack of separation was already noticed by \cite{salveson_1988}.
    On the other hand, when using subsetoids/PERs subtypes are trivially achieved by restricting both sides of the relation by the desired property.
    This leaves the representing type of the computational data unmodified: a fact not true in the setoid and univalent settings.
    This also ensures that proofs of logical properties have no bearing on the construction of computational data, and not by a restriction on elimination of inductively defined strict propositions but by the very structure of the formalization.
\end{remark}

\begin{construction}[Sub-\P-Sets]\label{Construction:SubPSet}
    Given a \P-set, \(A\), and a predicate on the elements of the underlying type of \(A\), \(P : |A| \rightarrow \Omega\), where \(\Omega\) in the \coq{} formalization is \texttt{SProp}, one can form the sub-\P-set, \(A\!\!\restriction_P\), where self-related elements must also satisfy \(P\), by the following:
    \begin{itemize}
        \item \(|\,A\!\!\restriction_P| \triangleq |A|\); and
        \item \(a \sim_{A\restriction_P} a' \triangleq\) \begin{itemize}
            \item \(a \sim_A a' \quad \wedge\)
            \item \(\makebox[0pt][l]{\(P\, a\)}\hphantom{a \sim_A a'} \quad \wedge\)
            \item \(P\, a'\).
        \end{itemize}
    \end{itemize}
\end{construction}

\subsection{Basic \P-Categorical Definitions}
We give a number of basic definitions for \P-category theory.
These largely replicate the definitions found in \E-category theory and standard mathematical accounts of category theory; although, there are a few deviations to account for the PER setting used for \P-category theory.

\begin{definition}[\texttt{PCat}]
    The definition of \emph{\P-categories}, found in listing~\ref{Listing:PCat}, is much the same as standard definitions found in the literature.
\end{definition}
There are a few key differences due to the \P-setting which we bring to the attention of the reader:
    \begin{itemize}
        \item the type of homs is valued in \texttt{PType};
        \item the identity arrow must be self-related; and
        \item the associativity and unit laws are phrased logically so as not to assume reflexivity.
    \end{itemize}
    Note that the latter two are particular to the \P-setting; \emph{i.e.}, they have no analogue in the \E-setting.
    \begin{listing}
        \begin{minted}[tabsize=2,breaklines,frame=lines]{coq}
Record PCat := {
  PCat_obj :> Type;
  PCat_hom : PCat_obj -> PCat_obj -> PType;

  PCat_id_mor : forall x, PCat_hom x x;
  PCat_comp : forall {x y z},
    PCat_hom y z -> PCat_hom x y -> PCat_hom x z;

  PCat_id_rel : forall x, PCat_id_mor x ~ PCat_id_mor x;

  PCat_comp_rel : forall {x y z}
    {f f' : PCat_hom y z} {g g' : PCat_hom x y},
    f ~ f' -> g ~ g' ->
    PCat_comp f g ~ PCat_comp f' g';

  PCat_assoc : forall {w x y z}
    {f f' : PCat_hom y z} {g g' : PCat_hom x y} {h h' : PCat_hom w x},
    f ~ f' -> g ~ g' -> h ~ h' ->
    PCat_comp (PCat_comp f g) h ~ PCat_comp f' (PCat_comp g' h');

  PCat_left_id : forall {x y} {f f' : PCat_hom x y},
    f ~ f' -> PCat_comp (PCat_id_mor y) f ~ f';

  PCat_right_id : forall {x y} {f f' : PCat_hom x y},
    f ~ f' -> PCat_comp f (PCat_id_mor x) ~ f';
}.
        \end{minted}
        \caption{\P-Category \coq{} Definition}
        \label{Listing:PCat}
    \end{listing}

\begin{construction}[Opposite \P-Category]
    The definition of a \P-category admits the standard construction of taking the opposite of a category.
\end{construction}

\begin{construction}[Product \P-Category]
    The definition of a \P-category admits the standard construction of taking the product of two categories.
\end{construction}

\begin{construction}[\texttt{PSet}]
    The \P-category, \texttt{PSet}, has as objects \texttt{PType}s, and has as homs the \texttt{PArr} of the domain and codomain \texttt{PType}s.
\end{construction}

\begin{definition}[\texttt{IsPIso}]
    A morphism, \(f : c \rightarrow d\), is a \emph{\P-isomorphism} when there exists another morphism, \(f^{-1} : d \rightarrow c\), such that:
    \begin{itemize}
        \item \(f \sim f\);
        \item \(f^{-1} \sim f^{-1}\);
        \item \(f \circ f^{-1} \sim \mathrm{id}_d\); and
        \item \(f^{-1} \circ f \sim \mathrm{id}_c\).
    \end{itemize}
    The \coq{} definition can be found in listing~\ref{Listing:PIso}.

    \begin{collect}{rocq}{}{}
        \begin{listing}[!htb]
\begin{minted}{coq}
Record IsPIso {C} {x y : C} (f : PCat_hom x y) := {
  IsPIso_rel : f ~ f;
  IsPIso_inv : PCat_hom y x;
  IsPIso_inv_rel : IsPIso_inv ~ IsPIso_inv;
  IsPIso_left : PCat_comp IsPIso_inv f ~ PCat_id_mor _; 
  IsPIso_right : PCat_comp f IsPIso_inv ~ PCat_id_mor _; 
}.
\end{minted}
            \caption{\P-Isomorphism \coq{} Definition}
            \label{Listing:PIso}
        \end{listing}
    \end{collect}
\end{definition}

\begin{definition}[\texttt{PFun}]
    The definition of \emph{\P-functors}, found in listing~\ref{Listing:PFun}, proceeds almost as straightforwardly as the definition of a \P-category.
\end{definition}
Much like the associativity and unit laws for \P-categories, the composition law for \P-functors requires extra hypotheses in order to account for the logical nature of the \P-setting.
    \begin{listing}
        \begin{minted}[tabsize=2,breaklines,frame=lines]{coq}
Record PFun (C D : PCat) := {
  PFun_obj_of :> C -> D;
  PFun_hom_of : forall {x y},
    PCat_hom x y -> PCat_hom (PFun_obj_of x) (PFun_obj_of y);

  PFun_hom_rel : forall {x y} {f f' : PCat_hom x y},
    f ~ f' -> PFun_hom_of f ~ PFun_hom_of f';

  PFun_comp_of : forall {x y z}
    {f f' : PCat_hom y z} {g g' : PCat_hom x y},
    f ~ f' -> g ~ g' ->
      PFun_hom_of (PCat_comp f g)
    ~
      PCat_comp (PFun_hom_of f') (PFun_hom_of g');

  PFun_id_of : forall x,
    PFun_hom_of (PCat_id_mor x) ~ PCat_id_mor (PFun_obj_of x);
}.
        \end{minted}
        \caption{\P-Functor \coq{} Definition}
        \label{Listing:PFun}
    \end{listing}

\begin{definition}
    \hfill
    \begin{itemize}
        \item (\P-Functor Fullness)
            A \P-functor, \(F : \mathbb{C} \rightarrow \mathbb{D}\), is \emph{\P-full} when it supports:
            \[ \textstyle
                \prod_{x\,y : \mathbb{C}}\prod_{f : F\,x \rightarrow F\,y}\,f \sim f \Rightarrow \sum_{g : x \rightarrow y}\, g \sim g \wedge F\,g \sim f \enspace.
            \]

        \item (\P-Functor Essential Surjectivity)
            A \P-functor, \(F : \mathbb{C} \rightarrow \mathbb{D}\), is \emph{\P-essentially surjective} when it supports:
            \[ \textstyle
                \prod_{d : \mathbb{D}}\sum_{c : \mathbb{C}}\, F\,c \cong d \enspace.
            \]
    \end{itemize}
    Note that we use a proof-relevant \textSigma-type rather than a proof-irrelevant mere existence quantification.
\end{definition}

\begin{construction}[\texttt{PHomFun}]
    Every \P-category, \(\mathbb{C}\), has a \P-functor mapping two objects to the \P-set of morphisms therebetween:
    \[ \mathrm{Hom}_{\mathbb{C}} : \mathbb{C}^{\mathsf{op}} \times \mathbb{C} \rightarrow \mathsf{PSet} \enspace. \]
\end{construction}

\begin{definition}[\texttt{IsPNatural}]
    A transformation,
    \[\textstyle \alpha : \prod_{c : \mathbb{C}} \mathrm{Hom}_\mathbb{D}(F \, c, G \, c) \enspace, \]
    is \P-natural precisely when:
    \[\textstyle \forall {x \, y :  \mathbb{C}}. \, \forall {f \, f' : x \rightarrow y}. \, f \sim f' \Rightarrow (\alpha_y \circ F \, f) \sim (G \, f' \circ \alpha_x) \enspace. \]
\end{definition}

\begin{definition}[\texttt{PNatTrans}]
    The definition of \emph{\P-natural transformations}, found in listing~\ref{Listing:PNatTrans}, is again a relatively mechanical translation of the ordinary definition into the \P-setting.
    \begin{listing}
        \begin{minted}[tabsize=2,breaklines,frame=lines]{coq}
Record PNatTrans {C D : PCat} (F G : PFun C D) := {
  PNatTrans_comps_of :> forall x, PCat_hom (F x) (G x);

  PNatTrans_comps_rel : forall x,
    PNatTrans_comps_of x ~ PNatTrans_comps_of x;

  PNatTrans_commutes : IsPNatural PNatTrans_comps_of;
}.
        \end{minted}
        \caption{\P-Natural Transformation \coq{} Definition}
        \label{Listing:PNatTrans}
    \end{listing}
\end{definition}

\subsection{\P-Functor Categories}
It would seem that we are now able to define easily the notion of \P-functor category: the objects are \P-functors and the morphisms are \P-natural transformations.
However, the situation for \P-functor categories is not as straightforward as the other \P-categorical definitions we have seen hitherto.
Indeed, here we find the first opportunity for the \P-setting to show its utility, rather than its encumbrances.
As with any \P-category, to define \P-functor categories
we have to consider how to relate the elements of the hom \P-sets.
Embracing partiality sheds new light on how to define \P-functor categories.

\begin{definition}[\P-Functor Categories]
     \P-Functor categories have \P-functors as objects.
     The hom \P-sets have \emph{all} transformations as the underlying type, but relating only those transformations that are \P-natural.

    Two morphisms, \(\alpha\) and \(\beta\), in a \P-functor category are related precisely when:
    \begin{itemize}
        \item \(\alpha\) and \(\beta\) are both \P-natural; and
        \item \(\alpha\) and \(\beta\) are componentwise related: \(\forall x. \, \alpha_x \sim \beta_x\).
    \end{itemize}
\end{definition}

\begin{remark}
    The above definition is the same as that given by \cite{cubric_normalization_1998}.
    We note that the first two conjuncts imply each other in the presence of the third one.
    Nevertheless, there is a simpler, and in our experience superior, equivalent definition.
    Transformations, \(\alpha\) and \(\beta\) of type \(F \Rightarrow G\), are related precisely when:
    \begin{itemize}
        \item \(\forall x\, y\, f\, f'. \, f \sim f' \Rightarrow \alpha_y \circ F \, f \sim G \, f' \circ \beta_x\); and
        \item \(\forall x. \, \alpha_x \sim \beta_x\)
    \end{itemize}
    This definition having fewer constituent parts simplifies the formalization and diminishes some of the inconvenience of using PERs, whilst maintaining the elegance of an unbiased symmetric definition.
\end{remark}

\begin{construction}[Presheaf \P-Categories]
    The presheaf \P-category, \(\widehat{\mathbb{C}}\), of a \P-category, \(\mathbb{C}\), is the \mbox{\P-functor} category, \([\mathbb{C}^{\mathsf{op}}, \mathsf{PSet}]\), from the opposite of \(\mathbb{C}\) to \texttt{PSet}.
\end{construction}

\begin{construction}[Sub-\P-Presheaves]\label{Construction:SubPPshf}
    Construction~\ref{Construction:SubPSet} can be extended to form sub-\P-presheaves.
    Given the following data:
    \begin{itemize}
        \item \(F : \widehat{\mathbb{C}}\); and
        \item \(P : \prod_{c : \mathbb{C}^{\mathsf{op}}}{F\,c \rightarrow \Omega}\); such that
        \item \(\forall\, x\, y\, f\, f'\, a\, a'. P_y\, a \wedge P_y\, a' \wedge a \sim_{F_y} a' \wedge f \sim f' \Rightarrow P_x\, (F_1\, f\, a) \wedge P_x\, (F_1\, f'\, a')\);
    \end{itemize}
    one may form the sub-\P-presheaf \(F\!\!\restriction_P : \widehat{\mathbb{C}}\) so that \((F\!\!\restriction_P)(c) \equiv (F\,c)\!\!\restriction_{P\,c}\).
\end{construction}

\begin{construction}[Nerve \P-Functors]
    Any \P-functor, \(F : \mathbb{C} \rightarrow \mathbb{D}\), induces a \P-functor:
    \[ \langle F \rangle : \mathbb{D} \rightarrow \widehat{\mathbb{C}} \enspace; \]
    where
    \[ \langle F \rangle(d)(c) \triangleq \mathrm{Hom}_\mathbb{D}(F(c), d) \enspace. \]
\end{construction}

\begin{construction}[Yoneda \P-Functor]
    The Yoneda \P-functor embeds a \P-category, \(\mathbb{C}\), into its \mbox{\P-category} of presheaves:
    \[ y : \mathbb{C} \rightarrow \widehat{\mathbb{C}} \enspace. \]
    It is the nerve of the identity \P-functor.
\end{construction}

\subsection{\P-Comma Categories}
We define comma \P-categories.
Their definition requires careful adaptation for the \P-setting.
Partiality has a subtle effect on the definition of comma \P-categories.
Since not all morphisms are self-related, one needs to ensure that the morphism contained within the objects of comma \P-categories is self-related.

\begin{definition}[Comma \P-Categories]
    The objects of the comma \P-category, \(F \downarrow G\), for \(F : \mathbb{B} \rightarrow \mathbb{D}\) and \(G : \mathbb{C} \rightarrow \mathbb{D}\), contain the following:
    \begin{itemize}
        \item \(b : \mathbb{B}\);
        \item \(c : \mathbb{C}\);
        \item \(f : F\,b \rightarrow G\,c\); such that
        \item \(f \sim f\).
    \end{itemize}
    The underlying type of the morphisms of the comma \P-category, \((b_1, c_1, f_1) \rightarrow (b_2, c_2, f_2)\), contains the following:
    \begin{itemize}
        \item \(g : b_1 \rightarrow b_2\); and
        \item \(h : c_1 \rightarrow c_2\).
    \end{itemize}
    The PER on these morphisms relates \((g, h) \sim (g', h')\) precisely when the following holds:
    \begin{itemize}
        \item \(g \sim g' : b_1 \rightarrow b_2\);
        \item \(h \sim h' : c_1 \rightarrow c_2\); and
        \item \(f_2 \circ F(g) \sim G(h') \circ f_1 : F\,b_1 \rightarrow G\,c_2\).
    \end{itemize}
\end{definition}

\begin{construction}[Comma \P-Category Projection \P-Functors]
    Comma \P-categories, \(F \downarrow G\), for \(F : \mathbb{B} \rightarrow \mathbb{D}\) and \(G : \mathbb{C} \rightarrow \mathbb{D}\), have two projection \P-functors:
    \begin{itemize}
        \item \(\mathrm{Dom} : F \downarrow G \rightarrow \mathbb{B}\); and
        \item \(\mathrm{Cod} : F \downarrow G \rightarrow \mathbb{C}\).
    \end{itemize}
\end{construction}

\begin{construction}[Induced \P-Functors into \P-Comma Categories]\label{Construction:CommaIndPFun}
    One may induce a \P-functor from a \P-category, \(\mathbb{B}\), into a \P-comma category, \((F \downarrow G)\), where \(F : \mathbb{C} \rightarrow \mathbb{E}\) and \(G : \mathbb{D} \rightarrow \mathbb{E}\), by the following data: \begin{itemize}
        \item a \P-functor, \(H : \mathbb{B} \rightarrow \mathbb{C}\);
        \item a \P-functor, \(K : \mathbb{B} \rightarrow \mathbb{D}\); and
        \item a \P-natural transformation, \(\alpha : F \circ H \Rightarrow G \circ K\).
    \end{itemize}
    We denote such induced \P-functors by \(\langle H \downarrow_\alpha K \rangle\); we may drop the subscript \(\alpha\) if it is evident from the context.
\end{construction}

\subsection{\P-Set-Valued (Co)Ends}
\label{Section:PSetsCoEnd}
We define how to construct (\P-)ends and (\P-)coends over the \P-category of \P-sets.
The \P-setting is crucial for producing definitions that will exhibit good computational behaviour.

\begin{definition}[\P-Set-Valued End]
    The \emph{end} of a \P-functor, \(F : \mathbb{C}^{\mathsf{op}} \times \mathbb{C} \rightarrow \mathsf{PSet}\), is given by \begin{itemize}
        \item \(\left| \int_{c : \mathbb{C}}{ F (c, c) } \right| \triangleq \prod_{c : \mathbb{C}}{ F (c, c) }\); and
        \item \(w \sim w' \triangleq\) \begin{itemize}
            \item \(\forall {x\, y}.\, \forall {f\, f' : x \rightarrow y}.\,{ f \sim f' \Rightarrow F \, (f, \mathsf{id}) \, (w\hphantom{{}'} \, y) \sim  F \, (\mathsf{id}, f') \, (w\hphantom{{}'} \, x)} \quad \wedge\)
            \item \(\forall {x\, y}.\, \forall {f\, f' : x \rightarrow y}.\,{ f \sim f' \Rightarrow F \, (f, \mathsf{id}) \, (w' \, y) \sim  F \, (\mathsf{id}, f') \, (w' \, x)} \quad \wedge\)
            \item \(\forall {z}.\,{w \, z \sim w' \, z}\).
        \end{itemize}
    \end{itemize}
\end{definition}

\begin{remark}
    The definition of ends for \P-sets uses well the partiality afforded by PERs.
    Indeed, it was in defining ends for \P-sets that we first began to appreciate the importance of using partiality.
    The first two conjuncts in the definition of the PER specify that the two wedges should be dinatural, and the third conjunct requires that the two wedges agree up to the relevant PER.
    Similarly to the definition of the PER for \P-functor category homs, note that the first two conjuncts imply each other in the presence of the third one.
\end{remark}

\begin{definition}[\P-Set-Valued Coend]
    The \emph{coend} of a \P-functor, \(F : \mathbb{C}^{\mathsf{op}} \times \mathbb{C} \rightarrow \mathsf{PSet}\), is given by \begin{itemize}
        \item \(\left| \int^{c : \mathbb{C}}{ F (c, c) } \right| \triangleq \sum_{c : \mathbb{C}}{ F (c, c) }\); and
        \item \(w \sim w'\) is inductively generated by the following: \begin{itemize}
            \item \(\forall {z}.\, \forall {s\, s' : F \, (z, z)}.\,{ s \sim s' \Rightarrow (z; s) \sim (z; s') }\);
            \item \(\forall {x\, y}.\, \forall {f\, f' : y \rightarrow x}.\, \forall {s\, s' : F \, (x, y)}.\,{f \sim f' \Rightarrow s \sim s' \Rightarrow (y; F\,(f, \mathsf{id}) \, s) \sim (x; F\,(\mathsf{id}, f') \, s')}\);
            \item \(\forall {x\, y}.\, \forall {f\, f' : y \rightarrow x}.\, \forall {s\, s' : F \, (x, y)}.\,{f \sim f' \Rightarrow s \sim s' \Rightarrow (x; F\,(\mathsf{id}, f) \, s) \sim (y; F\,(f', \mathsf{id}) \, s')}\); and
            \item \(w_1 \sim w_2 \wedge w_2 \sim w_3 \Rightarrow w_1 \sim w_3\).
        \end{itemize}
    \end{itemize}
\end{definition}

\begin{remark}
    The definition of ends and coends for \P-sets emphasizes the advantage that \P-sets have over \E-sets, and even QITs in univalent foundations: they leave the underlying computational structure unmodified.
    They therefore offer a cleaner setting for distinguishing computational data from logical properties.
    We believe that such a clean distinction is likely to have advantages for extraction of algorithms in proof-relevant settings.
\end{remark}

\section{\P-Categorical Structure}
\label{Section:PCategoricalStructure}
We now define \P-categorical structure required for describing simply typed syntax \P-categorically.
We cover this in three subsections: the definition of finite limits, the definition of Cartesian-closed structure, and the definition of Cartesian-pre-closed structure.
This latter structure is a novel definition needed for our \P-categorical analysis of unquotiented STLC syntax.

\cite{cubric_normalization_1998} define the categorical notions of \P-terminal object, \P-Cartesian product, and \P-Cartesian exponential in an equational style: they specify the object-forming and morphism-forming operations and the \P-equations that they should satisfy.
We adopt a different approach by only asking for an object-forming operation together with an adjointness property that it should satisfy.
This category-theoretic approach gives further confidence in the correctness of the \P-categorical definition; indeed, we recover all of the properties given by \cite{cubric_normalization_1998} thus connecting their definition with ours.

\subsection{Finite \P-Limits}
We define the following limits in \P-category theory, thus defining all finite limits: terminal objects, Cartesian products, and equalizers.

\begin{definition}[Terminal Object]
    A \emph{terminal object} in a \P-category is an object \(\top\) such that the following \P-isomorphism is \P-natural in \(x\).
    \[ \mathrm{Hom}_{\mathbb{C}}(x, \top) \cong \Delta_{\texttt{PUnit}} \]
    That is, the \P-set of morphisms into the terminal object is a terminal \P-set (up to \P-equivalence).
    This is straightforwardly phrased in \coq{} in listing~\ref{Listing:TermObj}, where \texttt{PBiFunPartialRight} constructs a \P-functor from a \P-functor out of a product \P-category by fixing the second argument to be the specified object, \texttt{PConstFun} produces a constant \P-functor out of the terminal category, and \texttt{PTermFun} is the unique \P-functor into the terminal category.
    \begin{listing}
        \begin{minted}[tabsize=2,breaklines,frame=lines]{coq}
Definition IsPTermObj {C : PCat} (term : C) :=
  PNatIso
    (PBiFunPartialRight PHomFun term)
    (PCompFun (PConstFun PUnit) PTermFun).
        \end{minted}
        \caption{\P-Terminal Object \coq{} Definition}
        \label{Listing:TermObj}
    \end{listing}
\end{definition}

\begin{definition}[Cartesian Product]
    A \P-category has \emph{Cartesian products} when it has a binary operation on objects \(- \, \times =\) such that the following family of \P-isomorphisms are \P-natural in \(x\).
    \[ \mathrm{Hom}_{\mathbb{C}}(x, a \times b) \cong \mathrm{Hom}_{\mathbb{C}\times\mathbb{C}}((x, x), (a, b)) : \langle - , = \rangle \]
    Moreover, we denote the projection maps as follows:
    \begin{align*}
         \pi_1 : a \times b \rightarrow a
         \enspace, \quad
         \pi_2 : a \times b \rightarrow b\enspace.
    \end{align*}
    This is translated into \coq{} in listing~\ref{Listing:CartProd}.
    \begin{listing}
        \begin{minted}[tabsize=2,breaklines,frame=lines]{coq}
Definition IsPCartProd {C : PCat} (prod : C -> C -> C) := forall a b,
  PNatIso
    (PBiFunPartialRight PHomFun (prod a b))
    (PCompFun
      (PBiFunPartialRight PHomFun (a, b))
      (POppFun (PPairFun PIdFun PIdFun))
    ).
        \end{minted}
        \caption{\P-Cartesian Product \coq{} Definition}
        \label{Listing:CartProd}
    \end{listing}
\end{definition}

\begin{definition}[Cartesian \P-Category]
    A \P-category is a \emph{Cartesian \P-category} when it has both a terminal object and Cartesian products.
    This is rendered in \coq{} in listing~\ref{Listing:CartCat}.
    \begin{collect}{rocq}{}{}
        \begin{listing}
\begin{minted}{coq}
Record PCartCat := {
  PCartCat_cat :> PCat; 

  PCC_term : PCartCat_cat;
  PCC_prod : PCartCat_cat -> PCartCat_cat -> PCartCat_cat;

  PCC_term_prop : IsPTermObj PCC_term;
  PCC_prod_prop : IsPCartProd PCC_prod;
}.
\end{minted}
            \caption{Cartesian \P-Category \coq{} Definition}
            \label{Listing:CartCat}
        \end{listing}
    \end{collect}
\end{definition}

\begin{definition}[Cartesian \P-Functor]
    A \P-functor between Cartesian \P-categories is a \emph{Cartesian \mbox{\P-functor}} when it preserves, up to \P-isomorphism, both the terminal object and the Cartesian products.
    We denote the inverse to the canonical map for products by the following:
    \[
        p : \left( F\,a \times F\,b \right)
        \stackrel{\raisebox{2mm}{\scriptsize$\sim$}}{\biarrow}
        F(a \times b) : \langle F\,\pi_1, F\,\pi_2 \rangle \enspace.
    \]
    This is rendered in \coq{} in listing~\ref{Listing:CartFun}.
    \begin{collect}{rocq}{}{}
        \begin{listing}
\begin{minted}{coq}
Record PCartFun (C D : PCartCat) := {
  PCartFun_fun :> PFun C D;

  PCartFun_term_of_iso :
    IsPIso (PCC_zero_mor (PCartFun_fun (PCC_term _)));

  PCartFun_prod_of_iso : forall a b,
    IsPIso
      (PCC_pair
        (PFun_hom_of PCartFun_fun (PCC_pi_left a b))
        (PFun_hom_of PCartFun_fun (PCC_pi_right a b))
      );
}.
\end{minted}
            \caption{Cartesian \P-Functor \coq{} Definition}
            \label{Listing:CartFun}
        \end{listing}
    \end{collect}
\end{definition}

\begin{lemma}[Cartesian Comma \P-Categories]
    Comma \P-categories, \((F \downarrow G)\), where \(F : \mathbb{B} \rightarrow \mathbb{D}\) and \(G : \mathbb{C} \rightarrow \mathbb{D}\), are \P-Cartesian whenever \(\mathbb{B}\), \(\mathbb{C}\), \(\mathbb{D}\) are Cartesian \P-categories, and \(G\) is a Cartesian \P-functor.
\end{lemma}

\begin{definition}[\P-Equalizers]
    A \P-category has \emph{\P-equalizers} when it has an operation
    \[ \textstyle
        \mathrm{eq}\{-, =\} : \prod_{a\, b : \mathbb{C}} \prod_{f\, g : a \rightarrow b} f \sim f \rightarrow g \sim g \rightarrow \mathbb{C}
    \]
    such that the following family of \P-isomorphisms are \P-natural in \(x\).
    \[ \textstyle
        \mathrm{Hom}_{\mathbb{C}}(x, \mathrm{eq}\{f, g\})
        \cong
        \{h : \mathrm{Hom}_{\mathbb{C}}(x, a) \,|\, f \circ h \sim g \circ h\}
    \]
    The \P-functor defined on the right is a sub-\P-presheaf of the \P-presheaf \(y(a)\).
    We therefore use the machinery of construction~\ref{Construction:SubPPshf} for defining sub-\P-presheaves to define this \P-functor.
    This definition is rendered in \coq{} in listing~\ref{Listing:Eq}, where \texttt{PSubSetFun} implements construction~\ref{Construction:SubPPshf} and requires an argument establishing the well-definedness of the subset predicate which we omit.
    \begin{listing}
        \begin{minted}[tabsize=2,breaklines,frame=lines]{coq}
Definition IsPEqualizer {C : PCat}
  (eq : forall (a b : C) (f g : PCat_hom a b), f ~ f -> g ~ g -> C) :=
    forall a b (f g : PCat_hom a b) (X1 : f ~ f) (X2 : g ~ g),
      PNatIso
        (PBiFunPartialRight PHomFun (eq a b f g X1 X2))
        (PSubSetFun
          (PBiFunPartialRight PHomFun a)
          (fun c (h : PCat_hom c a) => PCat_comp f h ~ PCat_comp g h)
          ( ... ) (* Proof that the subset predicate is well-defined. *)
        ).
        \end{minted}
        \caption{\P-Equalizers \coq{} Definition}
        \label{Listing:Eq}
    \end{listing}
\end{definition}

\begin{remark}
    Note that the object-forming operation for \P-equalizers requires that the two \P-morphisms are self-related.
    This is required so that the operation extends to a \P-functor out of a certain \P-comma category.
    Since we use \texttt{SProp} in our \coq{} formalization these proofs of self-relation have no computational content, and so we omit them by abuse of notation.
\end{remark}

\begin{definition}[Lex \P-Category]
    A Cartesian \P-category is a \emph{lex \P-category} when it has equalizers.
\end{definition}

\subsection{\P-Cartesian-Closed Structure}
We define Cartesian-closure in the \P-categorical setting.

\begin{definition}[Cartesian Exponential]
    A Cartesian \P-category has \emph{Cartesian exponentials} when it has a binary operation on objects \((-)^{(=)}\) such that the following family of \P-isomorphisms are \P-natural in \(x\).
    \[ \mathrm{Hom}_{\mathbb{C}}(x, b^a) \cong \mathrm{Hom}_{\mathbb{C}}(x \times a, b) : (-)^\ast \]
    The notation \(a \Rightarrow b\) may also be used instead of \(b^a\).
    Moreover, we denote the evaluation map as follows:
    \[ \varepsilon : b^a \times a \rightarrow b \enspace. \]
    This definition is translated into \coq{} in listing~\ref{Listing:CartExp}.
    \begin{listing}
        \begin{minted}[tabsize=2,breaklines,frame=lines]{coq}
Definition IsPCartExp {C : PCartCat} (exp : C -> C -> C) := forall a b,
  PNatIso
    (PBiFunPartialRight PHomFun (exp b a))
    (PCompFun
      PHomFun
      (PPairFun
        (POppFun (PBiFunPartialRight PCartProdFun a))
        (PCompFun (PConstFun b) PTermFun)
      )
    ).
        \end{minted}
        \caption{\P-Cartesian Exponential \coq{} Definition}
        \label{Listing:CartExp}
    \end{listing}
\end{definition}

\begin{definition}[Cartesian-Closed \P-Category]
    A Cartesian \P-category is a \emph{Cartesian-closed \P-cate\-gory} when it has Cartesian exponentials.
    This is rendered in \coq{} in listing~\ref{Listing:CartClosCat}.
    \begin{collect}{rocq}{}{}
        \begin{listing}
\begin{minted}{coq}
Record PCartClosCat := {
  PCartClosCat_cat :> PCartCat;

  PCCC_exp : PCartClosCat_cat -> PCartClosCat_cat -> PCartClosCat_cat;

  PCCC_exp_prop : IsPCartExp PCCC_exp;
}.
\end{minted}
            \caption{Cartesian-Closed \P-Category \coq{} Definition}
            \label{Listing:CartClosCat}
        \end{listing}
    \end{collect}
\end{definition}

\begin{definition}[Lex-Closed \P-Category]
    A Cartesian \P-category is a \emph{lex-closed \P-category} when it has equalizers and Cartesian exponentials.
\end{definition}

\begin{lemma}[\texttt{PSet} Cartesian Closure]
    The \P-category \texttt{PSet} is Cartesian-closed with the terminal object being given by \texttt{PUnit}, with the Cartesian product being given by \texttt{PProd}, and with the Cartesian exponential being given by \texttt{PArr}.
\end{lemma}

\begin{lemma}[\P-Presheaves Cartesian Closure]
    %\label{Construction:presheaf_cc}
    The \P-category of presheaves is Cartesian-closed with the terminal object being the constant presheaf on \texttt{PUnit}, and the Cartesian product being given pointwise.
    The Cartesian exponential in presheaves is involved but standard.
\end{lemma}

\begin{definition}[Cartesian-Closed \P-Functor]
    A Cartesian \P-functor between Cartesian-closed \P-cate\-gories is a \emph{Cartesian-closed \P-functor} when it preserves, up to \P-isomorphism, the Cartesian exponentials.
    We denote the inverse to the canonical map by the following:
    \[
        e : \left( F\,a \Rightarrow F\,b \right)
            \stackrel{\raisebox{2mm}{\scriptsize$\sim$}}{\biarrow}
        F(a \Rightarrow b) : (F\,\varepsilon \circ p)^\ast \enspace.
    \]
    This is rendered in \coq{} in listing~\ref{Listing:CartClosFun}.
    \begin{collect}{rocq}{}{}
        \begin{listing}
\begin{minted}{coq}
Record PCartClosFun (C D : PCartClosCat) := {
  PCartClosFun_fun :> PCartFun C D;

  PCartClosFun_exp_of_iso : forall a b,
    IsPIso
      (PCCC_transp
        (PCat_comp
          (PFun_hom_of PCartClosFun_fun (PCCC_eval a b))
          (PCartFun_prod_of _ a)
        )
      );
}.
\end{minted}
            \caption{Cartesian-Closed \P-Functor \coq{} Definition}
            \label{Listing:CartClosFun}
        \end{listing}
    \end{collect}
\end{definition}

\begin{lemma}[Nerve Cartesian-Closed \P-Functor]
    Any nerve \P-functor from a Cartesian \P-category is a Cartesian \P-functor, and the nerve of any Cartesian, essentially surjective, and full \P-functor is a Cartesian-closed \P-functor.
    In particular, the Yoneda \P-functor is a Cartesian-closed \P-functor.
\end{lemma}

\begin{lemma}[Cartesian-Closed Gluing \P-Categories]
    Gluing \P-categories, which are comma \P-cate\-gories \((\mathrm{Id}_\mathbb{D} \downarrow F)\) for \(F : \mathbb{C} \rightarrow \mathbb{D}\), are Cartesian-closed whenever \(\mathbb{C}\) is a Cartesian-closed \P-category, \(\mathbb{D}\) is a lex-closed \P-category, and \(F\) is a Cartesian \P-functor.
\end{lemma}

\subsection{\P-Cartesian-Pre-Closed Structure}
We define Cartesian-pre-closure in the \P-categorical setting.
This definition accounts for the structure of unquotiented syntax with its lack of equations for exponential-like structure.
We give a number of constructions, based on Cartesian-pre-closed structure, that we shall need for our normalization algorithms.

\begin{definition}[Cartesian Pre-Exponential]
    A Cartesian \P-category has \emph{Cartesian pre-exponentials} when it has a binary operation on objects \((-)^{(=)}\) together with a chosen pair of maps \P-natural in \(x\) as follows:
    \[ \mathrm{Hom}_{\mathbb{C}}(x, b^a) \mathrel{\biarrow} \mathrm{Hom}_{\mathbb{C}}(x \times a, b) : (-)^\ast \enspace. \]
    The notation \(a \Rightarrow b\) may also be used instead of \(b^a\).
    Moreover, we denote the pre-evaluation map as follows:
    \[ \tilde{\varepsilon} : b^a \times a \rightarrow b \enspace. \]
\end{definition}

\begin{remark}
    The definition of Cartesian pre-exponentials is an extreme weakening of the definition of Cartesian exponentials that does not require the two maps to be inverses of each other.
    This allows one to replicate the object-forming and morphism-forming structure of Cartesian exponentials in the absence of any of the equations establishing the \textbeta-laws and \texteta-laws.
\end{remark}

\begin{definition}[Cartesian-Pre-Closed \P-Category]
    A Cartesian \P-category is a \emph{Cartesian-pre-closed \P-category} when it has Cartesian pre-exponentials.
\end{definition}

\begin{definition}[Cartesian-Pre-Closed \P-Functor]
    A Cartesian \P-functor from a Cartesian-pre-closed \P-category to a Cartesian-closed \P-category is a \emph{Cartesian-pre-closed \P-functor} when it has the following structure and properties:
    \begin{itemize}
        \item \(\tilde{e} : (F\,b)^{(F\,a)} \rightarrow F\,(b^a)\); such that
        \item \(\tilde{e} \sim \tilde{e}\); and
        \item
        \(
        (f \sim g : x \times a \to b)
        \Rightarrow
        (\tilde{e} \circ (F\,f \circ p)^\ast \sim F\,(g^\ast) : F\,x\to F\,(b^a))
        \).
    \end{itemize}
\end{definition}

\begin{remark}
    The notion of Cartesian-pre-closed \P-functor is an appropriate weakening of the notion of Cartesian-closed \P-functor to the setting of having a Cartesian pre-closed domain category.
\end{remark}

Note that the definition of Cartesian-pre-closed \P-functors is heterogeneous: the structure of the domain and codomain \P-categories are different.
They are therefore not composable.
Although we do have the following lemma.

\begin{lemma} \label{lemma:CompClosedFunPreClosedFun}
    The composition of a Cartesian-closed \P-functor after a Cartesian-pre-closed \P-functor is a Cartesian-pre-closed \P-functor.
\end{lemma}

\begin{lemma}[Nerve Cartesian-Pre-Closed \P-Functor]
    The nerve of any Cartesian and essentially surjective \P-functor is a Cartesian-pre-closed \P-functor.
    In particular, the Yoneda Cartesian \P-functor is a Cartesian-pre-closed \P-functor.
\end{lemma}

\begin{lemma}
    For \(\mathbb{B}\) a Cartesian-pre-closed \P-category, \(\mathbb{C}\) a Cartesian-closed \P-category, \(\mathbb{D}\) a lex-closed \P-category, \(H : \mathbb B \to \mathbb D\) and \(K : \mathbb{B} \rightarrow \mathbb{C}\) Cartesian-pre-closed \P-functors, and \(F : \mathbb{C} \rightarrow \mathbb{D}\) a Cartesian \P-functor, if \(\alpha : H \Rightarrow F \circ K\) lifts the Cartesian-pre-closure of \(H\) and \(K\) as in figure~\ref{Figure:LiftCartPreClosureCommaIndPFun} then the \P-functor \(\langle H \downarrow_\alpha K \rangle : \mathbb{B} \to (\mathrm{Id}_\mathbb{D} \downarrow F)\) induced following construction~\ref{Construction:CommaIndPFun} is Cartesian-pre-closed.

    \begin{figure}
        \centering
        \[
            \left(\begin{tikzcd}
            	H\,b_1 \\
            	{} \\
            	F(K\,b_1)
            	\arrow["\alpha_{b_1}"{description}, from=1-1, to=3-1]
            \end{tikzcd}\right)
            \Rightarrow
            \left(\begin{tikzcd}
            	H\,b_2 \\
            	{} \\
            	F(K\,b_2)
            	\arrow["\alpha_{b_2}"{description}, from=1-1, to=3-1]
            \end{tikzcd}\right)
            \begin{tikzcd}
            	  {} & \phantom{H\,b_0} & {} \\
            	{} & {} & {} \\
            	{} & \phantom{FK\,b_0} & {}
            	\arrow["\tilde{e}_H \circ \mathrm{Dom}(\varepsilon)^\ast", from=1-1, to=1-3, dotted]
                \arrow[from=2-1, to=2-3]
                \arrow["\tilde{e}_K"', from=3-1, to=3-3, dotted]
            \end{tikzcd}
             \left(\begin{tikzcd}
            	H\,(b_1 \Rightarrow b_2) \\
            	{} \\
            	F(K\,(b_1 \Rightarrow b_2))
            	\arrow["\alpha_{b_1 \Rightarrow b_2}"{description}, from=1-1, to=3-1]
            \end{tikzcd}\right)
        \]
        \caption{Lifting of Cartesian-Pre-Closure to Induced \P-Functors into Comma \P-Categories}
        \label{Figure:LiftCartPreClosureCommaIndPFun}
    \end{figure}
\end{lemma}

\section{Simply Typed Syntax}
\label{Section:SimplyTypedSyntax}

We formalize the syntax of STLC in the well-scoped and well-typed style afforded by inductive families in \coq{}.
There have been many expositions on the formalization of syntax in proof-assistants for simply typed calculi.
We mainly follow the work of \cite{benton_strongly_2012} by using renamings and substitutions.
In their work, they use functional representations for renamings and substitutions and are forced to assume the axiom of function extensionality in \coq{} -- forfeiting canonicity of their formalization -- to prove many properties about their representation.
An alternative approach to using a functional representation is to use explicit lists to represent renamings and substitutions.
This representation has the advantage that it permits inductive reasoning more straightforwardly, at the cost of complicating the definitions of composition and proof of associativity which the functional representation gives ``for free''.
We use the list representation for the sake of inductive reasoning; although, had we adopted the functional approach we would not have been required to assume function extensionality as working in the \P-setting allows us to choose our (partial) equivalence relation for which we could have selected extensional equality.

\subsection{Types and Syntax}

We work in the simply typed setting with a single base type.

\begin{definition}[\texttt{Ty}]
    The set of \emph{types} is given by the grammar
    \[
    T ::= \iota \mid T\to  T \enspace,
    \]
    and inductively defined in the \coq{} listing~\ref{Listing:Ty}.
    \begin{collect}{rocq}{}{}
        \begin{listing}
\begin{minted}{coq}
Inductive Ty : Set :=
  | Iota : Ty
  | Arr : Ty -> Ty -> Ty.
\end{minted}
            \caption{Simple Types \coq{} Definition}
            \label{Listing:Ty}
        \end{listing}
    \end{collect}
\end{definition}

Contexts are lists of types.
We denote the empty context as \(\bullet\), and context extension with a comma, \emph{e.g.}, \(\Gamma, T\).
Such extended contexts may also be surrounded with parentheses,  \emph{e.g.}, \((\Gamma, T)\).
We introduce contexts as an inductive type so that we can name the constructors in the \coq{} formalization.

\begin{definition}[\texttt{Ctxt}]
    \emph{Contexts} are finite sequences of types.
    The \coq{} definition is in listing~\ref{Listing:Ctxt}.
        \begin{collect}{rocq}{}{}
            \begin{listing}
\begin{minted}{coq}
Inductive Ctxt : Set :=
  | Nil : Ctxt
  | Snoc : Ctxt -> Ty -> Ctxt.
\end{minted}
                \caption{Context \coq{} Definition}
                \label{Listing:Ctxt}
            \end{listing}
        \end{collect}
\end{definition}
We abuse notation and embed the set of types into the set of contexts by the following shorthand:
\[
    (T) \triangleq (\bullet, T) \enspace.
\]

A typed variable in context is represented by a well-scoped and well-typed de Bruijn index.

\begin{definition}[\texttt{Idx}]
    We define the type of \emph{indices}, \emph{\texttt{Idx}}, as the inductive family in \coq{} given in listing~\ref{Listing:Idx}.
    \begin{listing}
        \begin{minted}[tabsize=2,breaklines,frame=lines]{coq}
Inductive Idx : Ctxt -> Ty -> Set :=
  | IdxZero {Gamma T} : Idx (Snoc Gamma T) T
  | IdxSucc {Gamma T T'} : Idx Gamma T -> Idx (Snoc Gamma T') T.
        \end{minted}
        \caption{De Bruijn Index \coq{} Definition}
        \label{Listing:Idx}
    \end{listing}
\end{definition}

Well-scoped and well-typed STLC terms are given next.

\begin{definition}[\texttt{Tm}]
    We define the type of \emph{terms}, \emph{\texttt{Tm}}, as the inductive family in \coq{} in listing~\ref{Listing:Tm}.
    \begin{listing}
        \begin{minted}[tabsize=2,breaklines,frame=lines]{coq}
Inductive Tm Gamma : Ty -> Set :=
  | Var {T} : Idx Gamma T -> Tm Gamma T
  | App {T1 T2} : Tm Gamma (Arr T1 T2) -> Tm Gamma T1 -> Tm Gamma T2
  | Abs {T T'} : Tm (Snoc Gamma T) T' -> Tm Gamma (Arr T T').
        \end{minted}
        \caption{STLC Term \coq{} Definition}
        \label{Listing:Tm}
    \end{listing}
\end{definition}

\subsection{Renaming}
We equip our syntax of terms with a renaming structure: a substitution restricted to variables.
We emphasize, again, that we represent renamings with an explicit list rather than a type-preserving function from the variables in one context to the variables in another.

A context renaming is a list of variables taken from a given context.
The list of variables induces a second context given by its sequence of types.
We denote context renamings as arrows with a subscript \(\mathrm{ren}\), \emph{e.g.}, \(\rho : \Gamma \rightarrow_{\mathrm{ren}} \Delta\).

\begin{definition}[\texttt{CtxtRnm}]
    We define the type of \emph{context renamings}, \emph{\texttt{CtxtRnm}}, as the inductive family in \coq{} in listing~\ref{Listing:CtxtRnm}.
    \begin{listing}
        \begin{minted}[tabsize=2,breaklines,frame=lines]{coq}
Inductive CtxtRnm Gamma : Ctxt -> Set :=
  | CtxtRnmNil : CtxtRnm Gamma Nil
  | CtxtRnmSnoc {Delta T} : CtxtRnm Gamma Delta -> Idx Gamma T ->
    CtxtRnm Gamma (Snoc Delta T).
        \end{minted}
        \caption{Context Renaming \coq{} Definition}
        \label{Listing:CtxtRnm}
    \end{listing}
\end{definition}

\begin{construction}
    \hfill
    \begin{itemize}
        \item (Context Renaming Weakening)
            Any context renaming \(\rho: \Gamma \rightarrow_{\mathrm{ren}} \Delta\) may be weakened to \(\rho_T : (\Gamma, T) \rightarrow_{\mathrm{ren}} \Delta\).

        \item (Context Identity Renaming)
            For any context, \(\Gamma\), there is an identity renaming, \(\mathrm{id} : \Gamma \rightarrow_{\mathrm{ren}} \Gamma\).

        \item (Context Renaming Composition)
            Two context renamings, \(\rho : \Delta \rightarrow_{\mathrm{ren}} \Theta\) and \(\phi : {\Gamma \rightarrow_{\mathrm{ren}} \Delta}\), may be composed resulting in
            \[ (\rho \circ \phi) : \Gamma \rightarrow_{\mathrm{ren}} \Theta \enspace. \]
    \end{itemize}
    \end{construction}

\begin{remark}
    For all contexts, \(\Gamma\), and types, \(T\), there is a weakening renaming \((\mathrm{id})_T : (\Gamma,T) \to \Gamma\).
\end{remark}

\begin{lemma}
    Composing with the identity renaming is a neutral operation, and composition is associative.
\end{lemma}

\begin{construction}[Term Renaming]
    Any term, \(\Delta \vdash t : T\), may be renamed by \(\rho : \Gamma \rightarrow_{\mathrm{ren}} \Delta\) resulting in
    \[ \Gamma \vdash t[\rho] : T \enspace. \]
\end{construction}

\begin{theorem}[Term Renaming Action]
    Term renaming is an action: for all terms \(t\),
    \[ t[\mathrm{id}] = t \enspace, \quad t[\rho \circ \phi] = t[\rho][\phi] \enspace. \]
\end{theorem}

\subsection{Substitution}
We equip the syntax of terms with a substitution structure.
Again, we represent substitutions with an explicit list rather than a type-preserving function from the variables in one context to the terms in another.

A substitution is a list of terms taken from a given context.
The list of terms induces a second context given by its sequence of types.
We denote context substitutions as arrows with a subscript \(\mathrm{sub}\), \emph{e.g.}, \(\sigma : \Gamma \rightarrow_{\mathrm{sub}} \Delta\).

\begin{definition}[\texttt{CtxtSubst}]
    We define the type of \emph{context substitutions}, \emph{\texttt{CtxtSubst}}, as the inductive family in \coq{} in listing~\ref{Listing:CtxtSubst}.
    \begin{listing}
        \begin{minted}[tabsize=2,breaklines,frame=lines]{coq}
Inductive CtxtSubst Gamma : Ctxt -> Set :=
  | CtxtSubstNil : CtxtSubst Gamma Nil
  | CtxtSubstSnoc {Delta T} : CtxtSubst Gamma Delta -> Tm Gamma T ->
    CtxtSubst Gamma (Snoc Delta T).
        \end{minted}
        \caption{Context Substitution \coq{} Definition}
        \label{Listing:CtxtSubst}
    \end{listing}
\end{definition}

\begin{construction}
    \hfill{}
    \begin{itemize}
        \item (Context Substitution Weakening)
            Any context substitution \(\Gamma \rightarrow_{\mathrm{sub}} \Delta\) may be weakened so that it maps \((\Gamma, T) \rightarrow_{\mathrm{sub}} \Delta\).

        \item (Context Identity Substitution)
            For any context, \(\Gamma\), there is an identity substitution, \(\mathrm{id} : \Gamma \rightarrow_{\mathrm{sub}} \Gamma\).

        \item (Context Substitution Composition)
            Two context substitutions, \(\sigma : \Delta \rightarrow_{\mathrm{sub}} \Theta\) and \(\psi : \Gamma \rightarrow_{\mathrm{sub}} \Delta\), may be composed resulting in
            \[ (\sigma \circ \psi) : \Gamma \rightarrow_{\mathrm{sub}} \Theta \enspace. \]
    \end{itemize}
\end{construction}

\begin{lemma}
Composing with the identity substitution is a neutral operation, and composition is associative.
\end{lemma}

\begin{construction}[Term Substituting]
    Any term, \(\Delta \vdash t : T\), may be substituted using
    \(\sigma : \Gamma \rightarrow_{\mathrm{sub}} \Delta\) resulting in
    \[ \Gamma \vdash t[\sigma] : T \enspace. \]
\end{construction}

\begin{theorem}[Term Substituting Action]
Term substituting is an action: for all terms \(t\),
    \[ t[\mathrm{id}] = t \enspace, \quad t[\sigma \circ \psi] = t[\sigma][\psi] \enspace. \]
\end{theorem}

\begin{remark}
    Any context renaming may be transformed into an equivalent context substitution by a lifting operation.
    The lifting operation respects the identity, and composition of renamings.
    Moreover, the action of a lifted renaming on a term is the same as the action of the original renaming.
    This prevents any ambiguity in notation arising from, \emph{e.g.}, \(t[(\mathrm{id})_T]\) where it is ambiguous if the \((\mathrm{id})_T\) is a weakened renaming or a weakened substitution.
    Additionally, we formalize what may be referred to as \emph{``hetero-compositions''}: mixed compositions of renamings and substitutions.
    These hetero-compositions satisfy the appropriate unitality and correctness conditions.
    We also formalized some, but not all, of the ``associativity'' conditions for hetero-compositions as required by our development.
\end{remark}

\subsection{\textbeta\texteta-Conversion}
Hitherto we have introduced the syntax of STLC.
By using de Bruijn indices we have that \textalpha-equivalence of terms and of substitutions is just the standard identity strict proposition in \coq{}.
We now proceed to define \textbeta\texteta-conversion for both terms and substitutions.

\begin{definition}[\texttt{BetaEtaConv}]
    We adopt the definition of \emph{\textbeta\texteta-conversion} as the least symmetric, transitive, and congruent relation over \textbeta-reduction and \texteta-expansion.
    The \coq{} definition is given in listing~\ref{Listing:BetaEtaConv}.
    \begin{collect}{rocq}{}{}
        \begin{listing}
\begin{minted}{coq}
Inductive BetaEtaConv {Gamma} : forall {T},
  Tm Gamma T -> Tm Gamma T -> SProp :=
    | BetaEtaConvVar {T} i : @BetaEtaConv _ T (Var i) (Var i)
    | BetaEtaConvApp {T1 T2} {t1 t1'} {t2 t2'} :
      @BetaEtaConv _ (Arr T1 T2) t1 t1' ->
      @BetaEtaConv _ T1 t2 t2' ->
      BetaEtaConv (App t1 t2) (App t1' t2')
    | BetaEtaConvAbs {T T'} {t1 t2} :
      BetaEtaConv t1 t2 ->
      @BetaEtaConv _ (Arr T T') (Abs t1) (Abs t2)
    | BetaEtaConvBeta {T1 T2} t1 t2 :
      BetaEtaConv (App (Abs t1) t2) (@BetaSubst _ T1 T2 t1 t2)
    | BetaEtaConvEta {T1 T2} t :
      BetaEtaConv t (Abs (App (@Shift _ T1 T2 t) (Var IdxZero)))
    | BetaEtaConvSymm {T} {t1 t2} :
      BetaEtaConv t1 t2 ->
      @BetaEtaConv _ T t2 t1
    | BetaEtaConvTrans {T} {t1 t2 t3} :
      BetaEtaConv t1 t2 ->
      BetaEtaConv t2 t3 ->
      @BetaEtaConv _ T t1 t3.
\end{minted}
            \caption{Term \textbeta\texteta-Conversion \coq{} Definition}
            \label{Listing:BetaEtaConv}
        \end{listing}
    \end{collect}
    There, we use two helper functions: \texttt{BetaSubst} which substitutes the second argument for the de Bruijn index zero in the first argument thereby decreasing the length of the context; and \texttt{Shift} which introduces a new free variable at de Bruijn index zero thereby increasing the length of the context.
\end{definition}

\begin{remark}
    We omit a constructor for reflexivity opting to prove it as a corollary of variables being self-convertible -- which we include as part of the definition of being a congruence over the nullary (with respect to terms) variable constructor -- and \textbeta\texteta-conversion being a congruence.
\end{remark}

\begin{definition}[\texttt{CtxtSubstConv}]
    The extension of \emph{\textbeta\texteta-conversion} of terms to substitutions via a list construction is in the \coq{} code in listing~\ref{Listing:BetaEtaConvSubst}.
    \begin{collect}{rocq}{}{}
        \begin{listing}
\begin{minted}{coq}
Inductive CtxtSubstConv {Gamma} : forall {Delta},
  CtxtSubst Gamma Delta -> CtxtSubst Gamma Delta -> SProp :=
    | CtxtSubstConvNil : CtxtSubstConv (CtxtSubstNil _) (CtxtSubstNil _)
    | CtxtSubstConvSnoc {Delta s1 s2 T t1 t2} :
      @CtxtSubstConv Gamma Delta s1 s2 ->
      @BetaEtaConv _ T t1 t2 ->
      CtxtSubstConv (CtxtSubstSnoc s1 t1) (CtxtSubstSnoc s2 t2).
\end{minted}
            \caption{Context Substitution \textbeta\texteta-Conversion \coq{} Definition}
            \label{Listing:BetaEtaConvSubst}
        \end{listing}
    \end{collect}
\end{definition}

\section{Categorical Structure of Simply Typed Syntax}
\label{Section:CategoricalStructure}
We give a number of definitions of \P-categories and \P-functors that arise out of the structure of syntax.
Thereafter, we proceed to give these emergent structures universal characterizations firmly placing them within the context of (\P-)categorical analysis.

\subsection{Definitions}
We define three \P-categories that we shall use to examine the universal structure of syntax \P-categorically.
Each of these \P-categories will have contexts for their objects, but will have different families of hom \P-sets.

\begin{construction}[\texttt{Rnm}]
    The \P-category of contexts and context renamings \texttt{Rnm} has as objects contexts and as morphisms context renamings considered as a discrete \texttt{PType}.

    \texttt{Rnm} is a Cartesian \P-category with the terminal object being the empty context and the Cartesian product being context concatenation.
\end{construction}

\begin{remark}
    Although not required for our development and so left unformalized, \texttt{Rnm} is the free Cartesian \P-category over the set of types \texttt{Ty}; \emph{i.e.}, up to \P-isomorphism, there is a unique interpretation Cartesian \P-functor from \texttt{Rnm} into any other Cartesian \P-category given an interpretation of \texttt{Ty} in it.
    This is a simpler form of the result in theorem~\ref{Theorem:SubstBetaEtaFree}.
\end{remark}

\begin{construction}[\texttt{Subst}\textsubscript{\textalpha}]
    \label{Construction:SubstAlphaStructure}
    The \P-category of contexts and context substitutions, \texttt{Subst}\textsubscript{\textalpha}, has as objects contexts and as morphisms context substitutions related by \textalpha-equivalence.

    \texttt{Subst}\textsubscript{\textalpha} is a Cartesian-pre-closed \P-category with the terminal object being given by the empty context, and the Cartesian product being given by context concatenation, which we denote by \(\Gamma \mathbin{+\!\!+} \Delta\).
    The pre-exponential is given by the following recursive definitions:
    \begin{align}
        {\Gamma^T} &\triangleq {\begin{cases}
            ( \bullet ) & \Gamma = ( \bullet ) \\
            ( {\Gamma'}^T, T \rightarrow T' ) & \Gamma = ( \Gamma' , T' )
        \end{cases}} \label{Equation:CtxtTyPreExp} \\
        {\Gamma^\Delta} &\triangleq {\begin{cases}
            \Gamma & \Delta = ( \bullet ) \\
            \makebox[0pt][l]{$\left( \Gamma^T \right)^{\Delta'}$} \hphantom{( {\Gamma'}^T, T \rightarrow T' )} & \Delta = ( \Delta' , T ) \label{Equation:CtxtPreExp}
        \end{cases}}
    \end{align}
    where definition~\ref{Equation:CtxtTyPreExp} defines pre-exponentiation of a context by a type and definition~\ref{Equation:CtxtPreExp} defines context pre-exponentiation.
    The pre-abstraction and pre-evaluation maps,
    \[
        \mathsf{Subst}_\alpha\left(\Gamma \mathbin{+\!\!+} \Delta, \Theta\right) \mathrel{\biarrow}  \mathsf{Subst}_\alpha\left(\Gamma, \Theta^\Delta \right) \enspace.
    \]
    are defined by induction from the basic maps
    \[
        \mathsf{Subst}_\alpha\left((\Gamma, T), (S)\right) \mathrel{\biarrow}  \mathsf{Subst}_\alpha\left(\Gamma, (T \rightarrow S)\right)
    \]
    given by
    \begin{align*}
        t & \mapsto \mathsf{Abs}\,t \enspace \textrm{; and}
        \\
        \mathsf{App} \, \left(t[(\mathrm{id})_T]\right) \left(\mathsf{Var}\,\mathsf{IdxZero}\right)
        & \mapsfrom t \enspace.
    \end{align*}
    Both of these constructions follow a two-step process where they are first constructed with respect to pre-exponentiation by types, and thereafter constructed with respect to pre-exponentiation by contexts.
\end{construction}

\begin{construction}[\texttt{Subst}\textsubscript{\textbeta\texteta}]
    The \P-category of contexts and context substitutions, \texttt{Subst}\textsubscript{\textbeta\texteta}, has as objects contexts and as morphisms context substitutions related by \textbeta\texteta-conversion.

    \texttt{Subst}\textsubscript{\textbeta\texteta} is a Cartesian-closed \P-category with the terminal object being given by the empty context, and the Cartesian product being given by context concatenation.
    The Cartesian exponential is given as in construction~\ref{Construction:SubstAlphaStructure}.
\end{construction}

\begin{construction}[Inclusion and Quotient \P-Functors]
    We have the following identity-on-objects and Cartesian \P-functors:
    \begin{itemize}
        \item \(i : \mathsf{Rnm} \rightarrow \mathsf{Subst}_\alpha\), and
        \item \(j : \mathsf{Subst}_\alpha \rightarrow \mathsf{Subst}_{\beta\eta}\).
    \end{itemize}
    Moreover, the latter \P-functor is Cartesian-pre-closed and full.
\end{construction}

\subsection{Universal Characterizations}

\begin{figure}
    \centering
    \begin{tikzcd}
        {\mathcal{F}} \arrow[rr, start anchor=north east, end anchor=north west, bend left, ""{name=S, below}, "{\exists \llbracket - \rrbracket}"]
                      \arrow[rr, start anchor=south east, end anchor=south west, bend right, ""{name=I, above}, "{\forall I}"']
                      && \mathbb{C}
        \arrow[Rightarrow, from=S, to=I, shift right=3, "{\exists ! q}"']
        \arrow[Leftarrow, from=S, to=I, shift left=1, "{u = q^{-1}}"]
    \end{tikzcd}
    \vspace{2ex}
    \caption{Freeness Property for a Free Cartesian-Closed \P-Category}
    \label{Figure:FreeCCC}
\end{figure}
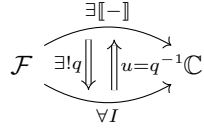

\begin{definition}[Free Cartesian-Closed \P-Category]\label{Definition:FreeCCC}
    A \P-category, \(\mathcal{F}\), is a \emph{free Cartesian-closed \P-cate\-gory over a base type} when it supports the following structure and properties: for all Cartesian-closed \mbox{\P-categories} \(\mathbb{C}\),
    \begin{enumerate}
        \item \( \forall \,
        (c : \mathbb{C}).\, \exists \, (\llbracket - \rrbracket_c : \mathcal{F} \rightarrow_{\mathrm{CC}} \, \mathbb{C}).\, \llbracket (\iota) \rrbracket_c \equiv c \);

        \item \( \forall \,
        (c\, c' : \mathbb{C})\, (f : c \cong c').\, \exists \, (\llbracket - \rrbracket_f : \llbracket - \rrbracket_c \cong \llbracket - \rrbracket_{c'}). \, \llbracket (\iota) \rrbracket_f \equiv f \);

        \item \( \forall \,
        (c : \mathbb{C}).\, \llbracket - \rrbracket_{\mathrm{id}_c} \sim \mathrm{id}_{\llbracket - \rrbracket_c}\);

        \item \( \forall \,
        (c\, c'\, c'' : \mathbb{C})\, (f : c \cong c') \, (g : c' \cong c'').\, \llbracket - \rrbracket_{g \circ f} \sim \llbracket - \rrbracket_g \circ \llbracket - \rrbracket_f \); and

        \item \( \forall \,
        (I : \mathcal{F} \rightarrow_{\mathrm{CC}} \mathbb{C}). \, \exists \, (q : \llbracket - \rrbracket_{I_0(\iota)} \cong I : u).\) \begin{itemize}
            \item \(q_{\left(\iota\right)} \equiv \mathrm{id} \qquad \wedge\)
            \item \(u_{\left(\iota\right)} \equiv \mathrm{id} \qquad \wedge\)
            \item \(\forall \, (\alpha : \llbracket - \rrbracket_{I_0(\iota)} \overset{\!\!\sim}{\Rightarrow} I : \beta). \, \alpha_{\left(\iota\right)} \sim \mathrm{id} \wedge \beta_{\left(\iota\right)} \sim \mathrm{id} \Rightarrow q \sim \alpha \wedge u \sim \beta \).
        \end{itemize}
    \end{enumerate}
    Where ``\(\rightarrow_\mathrm{CC}\!\)'' above means a Cartesian-closed \P-functor.
\end{definition}
The first condition provides existence of a Cartesian-closed interpretation \P-functor into any pointed Cartesian-closed \P-category.
The second, third, and fourth conditions ensure that the interpretation \P-functor is coherently invariant under \P-isomorphisms of the pointing of the target Cartesian-closed \P-category.
Finally, the fifth condition ensures that the interpretation \P-functor is unique up to unique \P-isomorphism.
This is informally summarized in figure~\ref{Figure:FreeCCC}

\begin{remark}
    Definition~\ref{Definition:FreeCCC} is equivalent with \(\mathcal{F}\) being bicategorically \P-initial in the locally-groupoidal \P-bicategory of pointed Cartesian-closed \P-categories, pointed Cartesian-closed \P-functors, and pointed \P-natural isomorphisms.
    We have however chosen to use the elementary phrasing to emphasize the r\^ole of arbitrary Cartesian-closed interpretations, \(I : \mathcal{F} \rightarrow_{\mathrm{CC}} \mathbb{C}\), in inducing specific interpretations, \(\llbracket - \rrbracket_{I_0(\iota)} : \mathcal{F} \rightarrow_{\mathrm{CC}} \, \mathbb{C}\), and the associated \(q\)/\(u\) \P-natural isomorphisms therebetween.
\end{remark}

\begin{theorem}
    \label{Theorem:SubstBetaEtaFree}
    \texttt{Subst}\textsubscript{\textbeta\texteta} is a free Cartesian-closed \P-category over a single base type.
\end{theorem}

Importantly, the \P-functor \(j : \mathsf{Subst}_\alpha \rightarrow \mathsf{Subst}_{\beta\eta}\) satisfies the two-dimensional universal property informally summarized in figure~\ref{Figure:RelativeUniversalProperty}.
    \begin{figure}
        \centering
        \begin{tikzcd}
            & {\mathsf{Rnm}}
                \arrow[rd, start anchor=south east, "{i}"{above}, ""{name=A, below}]
                \arrow[ld, start anchor=south west, "{i}"{above}] \\
        	{\mathsf{Subst}_\alpha}
                \arrow[d, "{j}"{left}]
            && {\mathsf{Subst}_\alpha}
                \arrow[d, "{\forall I}"{right}] \\
        	{\mathsf{Subst}_{\beta\eta}}
                \arrow[rr, "{\exists \llbracket - \rrbracket}"{below}]
                \arrow[Rightarrow, from=A, shift right=2, "{\exists ! \tilde{u} \quad}"{above}, shorten <=3ex, shorten >=3ex]
                \arrow[Leftarrow, from=A, shift left=2, "{\quad \exists ! \tilde{q}}"{below}, shorten <=3ex, shorten >=3ex]
            && {\mathbb{C}}
        \end{tikzcd}
        \vspace{2ex}
        \caption{Universal Property for \texttt{Subst}\textsubscript{\textalpha} relative to \texttt{Subst}\textsubscript{\textbeta\texteta}}
        \label{Figure:RelativeUniversalProperty}
    \end{figure}

\begin{theorem} \label{Theorem:RelativeUniversalProperty}
    The \P-functor \(j : \mathsf{Subst}_\alpha \rightarrow \mathsf{Subst}_{\beta\eta}\) satisfies the property that for all Cartesian-closed \P-categories, \(\mathbb{C}\), and for all Cartesian-pre-closed \P-functors, \(I : \mathsf{Subst}_\alpha \to \mathbb{C}\),
    \vskip 0.5\baselineskip
    \noindent
    \( \exists \, (\tilde{q} : \llbracket - \rrbracket_{I_0(\iota)} \circ j \circ i \Rightarrow I \circ i) \, (\tilde{u} : I \circ i \Rightarrow \llbracket - \rrbracket_{I_0(\iota)} \circ j \circ i). \)
    \begin{enumerate}
        \item \(\tilde{q}_{\left(\iota\right)} \equiv \mathrm{id} \qquad \wedge\)
        \item \(\tilde{u}_{\left(\iota\right)} \equiv \mathrm{id} \qquad \wedge\)
        \item \(\forall \, (\alpha : \llbracket - \rrbracket_{I_0(\iota)} \circ j \circ i \Rightarrow I \circ i) \, (\beta :  I \circ i \Rightarrow \llbracket - \rrbracket_{I_0(\iota)} \circ j \circ i).\) \hfill \\
            \hspace*{0.5em}
            \(
                \begin{aligned}
                    & \alpha_{\left(\iota\right)} \sim \mathrm{id} & \wedge \\
                    & \beta_{\left(\iota\right)} \sim \mathrm{id} & \wedge \\
                    & \forall \, \Gamma \, \Delta. \, \alpha_{\Gamma \Rightarrow \Delta} \circ e_{\llbracket - \rrbracket} \sim \tilde{e}_I \circ (\beta_\Gamma \Rightarrow \alpha_\Delta) & \wedge \\
                    & \forall \, \Gamma \, \Delta. \, (\llbracket \varepsilon \rrbracket_{I_0(\iota)} \circ p)^\ast \circ \beta_{\Gamma \Rightarrow \Delta} \sim (\alpha_\Gamma \Rightarrow \beta_\Delta) \circ (I_1(\tilde{\varepsilon}) \circ p)^\ast & \Rightarrow \\
                    & \tilde{q} \sim \alpha \wedge \tilde{u} \sim \beta & \\
                \end{aligned}
            \).
    \end{enumerate}
\end{theorem}
This property is a variation of the final property of definition~\ref{Definition:FreeCCC} to account for the Cartesian-pre-closed structure of \texttt{Subst}\textsubscript{\textalpha}.
The first and second subconditions describe the strictness of the \P-natural transformations \(\tilde{q}\) and \(\tilde{u}\) at the base type.
The third subcondition is a uniqueness condition of \(\tilde{q}\) and \(\tilde{u}\), and thus a quasi-uniqueness condition on \(\llbracket - \rrbracket_{I_0(\iota)}\) with respect to the Cartesian-pre-closed structure \(\tilde{e}_I\) of \(I\).

\section{Computational Application: Normalization by Evaluation}
\label{Section:ComputationalApplication}

We are finally in a position to state our \P-categorical normalization functions for simply typed \textlambda-calculus terms.

\subsection{Soundness}

As described in figure~\ref{Figure:Norm1}, there are two (\emph{intensionally different}) canonical \P-functors from \texttt{Subst}\textsubscript{\textbeta\texteta} into presheaves thereover: the Yoneda embedding and the interpretation \P-functor.
\begin{figure}
    \centering
    \begin{tikzcd}
        {\makebox[0pt][l]{$\mathsf{Subst}_{\beta\eta}$} \phantom{\widehat{\mathsf{Subst}_{\beta\eta}}}} && \widehat{\mathsf{Subst}}_{\beta\eta}
        \arrow[rightarrow, from=1-1, to=1-3, start anchor=north east, end anchor=north west, bend left, ""{name=S, below}, "{\llbracket - \rrbracket}"]
        \arrow[rightarrow, from=1-1, to=1-3, start anchor=south east, end anchor=south west, bend right, ""{name=I, above}, "{y}"']
        \arrow[Rightarrow, from=S, to=I, shift right=3, "{q}"']
        \arrow[Leftarrow, from=S, to=I, shift left=3, "{u}"]
    \end{tikzcd}
    \vspace{2ex}
    \caption{Interpretation of \texttt{Subst}\textsubscript{\textbeta\texteta} into its \P-Category of Presheaves}
    \label{Figure:Norm1}
\end{figure}
This induces \P-natural isomorphisms:
    \[ q : \llbracket - \rrbracket \Rightarrow y \textrm{ with inverse } u : y \Rightarrow \llbracket - \rrbracket \enspace. \]
These \P-isomorphisms establish that the Yoneda embedding and the interpretation \P-functor are \emph{extensionally the same}.
From these two perspectives one can normalize any context substitution.

\begin{definition}[\texttt{nf}\textsubscript{1}]
    Following~\cite{cubric_normalization_1998}, we define a normalization function thus:
    \[ \mathsf{nf}_1(\sigma : \Gamma \rightarrow_{\mathrm{sub}} \Delta) \triangleq q_{\Delta, \Gamma}(\llbracket \sigma \rrbracket_{\Gamma}(u_{\Gamma,\Gamma}(\mathrm{id}_{\Gamma}))) \enspace, \]
    based on figure~\ref{Figure:Norm1}.
\end{definition}

\begin{definition}[\texttt{nf}\textsubscript{2}]
    Alternatively, using the Cartesian-closure of \(\langle j \rangle\), one may define a normalization function thus:
    \[ \mathsf{nf}_2(\sigma : \Gamma \rightarrow_{\mathrm{sub}} \Delta) \triangleq q_{\Delta, \Gamma}(\llbracket \sigma \rrbracket_{\Gamma}(u_{\Gamma,\Gamma}(\mathrm{id}_{\Gamma}))) \enspace, \]
    based on figure~\ref{Figure:Norm2}.
\end{definition}

\begin{figure}
    \centering
    \begin{tikzcd}
        {\makebox[0pt][l]{$\mathsf{Subst}_{\beta\eta}$} \phantom{\widehat{\mathsf{Subst}_{\beta\eta}}}} && \widehat{\mathsf{Subst}}_{\alpha}
        \arrow[rightarrow, from=1-1, to=1-3, start anchor=north east, end anchor=north west, bend left, ""{name=S, below}, "{\llbracket - \rrbracket}"]
        \arrow[rightarrow, from=1-1, to=1-3, start anchor=south east, end anchor=south west, bend right, ""{name=I, above}, "{\langle j \rangle}"']
        \arrow[Rightarrow, from=S, to=I, shift right=3, "{q}"']
        \arrow[Leftarrow, from=S, to=I, shift left=3, "{u}"]
    \end{tikzcd}
    \vspace{2ex}
    \caption{Interpretation of \texttt{Subst}\textsubscript{\textbeta\texteta} into the \P-Category of Presheaves over \texttt{Subst}\textsubscript{\textalpha}}
    \label{Figure:Norm2}
\end{figure}

\begin{theorem}[Soundness]
    \label{Theorem:Soundness}
    The normalization functions \(\mathsf{nf}_1\) and \(\mathsf{nf}_2\) are sound.
\end{theorem}
\begin{proof}
    The result follows by equational reasoning: abstracting \(\mathsf{nf}_i\) as \(\mathsf{nf}\), and \(y\) and \(\langle j \rangle\) as \(I\), we have:
    \begin{align*}
        \mathsf{nf}(\sigma) &\equiv q_{\Delta, \Gamma}(\llbracket \sigma \rrbracket_{\Gamma}(u_{\Gamma,\Gamma}(\mathrm{id}_{\Gamma}))) & \mathrm{def} \\
        &\sim_{\beta\eta} q_{\Delta, \Gamma}(u_{\Delta,\Gamma}(I(\sigma)_{\Gamma}(\mathrm{id}_{\Gamma}))) & \mathrm{nat} \\
        &\sim_{\beta\eta} q_{\Delta, \Gamma}(u_{\Delta,\Gamma}(\sigma)) & \mathrm{yon/nerve} \\
        &\sim_{\beta\eta} \sigma \enspace. & \mathrm{iso}
    \end{align*}
\end{proof}

\begin{remark}
    The \coq{} formalization of theorem~\ref{Theorem:Soundness} gives a program for computing a derivation of the \textbeta\texteta-conversion of a term with its normal form as an element of the type in listing~\ref{Listing:BetaEtaConv}.
    We leave the analysis of these to further work.
\end{remark}

The normalization function and category theory formalized, provide the following.
\begin{corollary}[Weak Completeness]
    \label{Corollary:WeakCompleteness}
    The normalization functions \(\mathsf{nf}_1\) and \(\mathsf{nf}_2\) are weakly complete:
    \[
        \sigma \sim_{\beta\eta} \sigma'
        \ \Rightarrow \
        \mathsf{nf}_1(\sigma) \sim_{\beta\eta} \mathsf{nf}_1(\sigma')
        \,\wedge\,
        \mathsf{nf}_2(\sigma) \sim_{\beta\eta} \mathsf{nf}_2(\sigma') \enspace.
    \]
\end{corollary}
\begin{remark}
    Although corollary~\ref{Corollary:WeakCompleteness} is, by symmetry and transitivity of \(\sim_{\beta\eta}\), a straightforward consequence of theorem~\ref{Theorem:Soundness}, a more direct proof can be given by observing that the normalization functions \(\mathsf{nf}_1\) and \(\mathsf{nf}_2\) are morphisms in \texttt{PSet} and therefore respect the appropriate PERs.
\end{remark}

The normalization functions \(\mathsf{nf}_1\) and \(\mathsf{nf}_2\) do not obviously satisfy strong completeness.
Although, by example computations, one can observe that neither is the identity function intensionally, we have no proof that they are canonicalization functions reducing \textbeta\texteta-conversion to \textalpha-equivalence.
We overcome this shortcoming by using the universal property of \texttt{Subst}\textsubscript{\textalpha} relative to \texttt{Subst}\textsubscript{\textbeta\texteta}
to construct strongly complete normalization functions.

\subsection{Strong Completeness}
We use the structure of Cartesian-pre-closure to derive a strongly complete normalization algorithm.

\begin{definition}[\texttt{nf}\textsubscript{3}]
    Using the Cartesian-pre-closure of \(y\), we define a normalization function thus:
    \[ \mathsf{nf}_3(\sigma : \Gamma \rightarrow_{\mathrm{sub}} \Delta) \triangleq \tilde{q}_{\Delta, \Gamma}(\llbracket \sigma \rrbracket_{\Gamma}(\tilde{u}_{\Gamma,\Gamma}(\mathrm{id}_{\Gamma}))) \enspace, \]
    based on figure~\ref{Figure:Norm3}.

    \begin{figure}
        \centering
        \begin{tikzcd}
            & {\mathsf{Rnm}}
                \arrow[rd, start anchor=south east, "{i}"{above}, ""{name=A, below}]
                \arrow[ld, start anchor=south west, "{i}"{above}] \\
            {\mathsf{Subst}_\alpha}
                \arrow[d, "{j}"{left}]
            && {\mathsf{Subst}_\alpha}
                \arrow[d, "{y}"{right}] \\
            {\mathsf{Subst}_{\beta\eta}}
                \arrow[rr, "{\llbracket - \rrbracket}"{below}]
                \arrow[Rightarrow, from=A, shift right=2, "{\tilde{u}}"{above}, shorten <=3ex, shorten >=3ex]
                \arrow[Leftarrow, from=A, shift left=2, "{\tilde{q}}"{below}, shorten <=3ex, shorten >=3ex]
            && {\widehat{\mathsf{Subst}}_{\alpha}}
        \end{tikzcd}
        \vspace{2ex}
        \caption{Interpretation of \texttt{Subst}\textsubscript{\textbeta\texteta} into the \P-Category of Presheaves over \texttt{Subst}\textsubscript{\textalpha}}
        \label{Figure:Norm3}
    \end{figure}
\end{definition}

\begin{theorem}[Strong Completeness]
    \label{Theorem:StrongCompleteness3}
    The normalization function \(\mathsf{nf}_3\) is strongly complete:
    \[ \sigma \sim_{\beta\eta} \sigma' \Rightarrow \mathsf{nf}_3(\sigma) \equiv_\alpha \mathsf{nf}_3(\sigma') \enspace. \]
\end{theorem}

The normalization function \(\mathsf{nf}_3\) is strongly complete but it is not obviously sound as the \(\tilde{q}\) and \(\tilde{u}\) associated to the universal property of \texttt{Subst}\textsubscript{\textalpha} relative to \texttt{Subst}\textsubscript{\textbeta\texteta} are not \P-natural with respect to substitutions, but only renamings, and so our former proof of soundness cannot be translated across to the new setting.
Although, by example computations, one can observe that it seems to be the identity function extensionally, we have no proof that its output is \textbeta\texteta-convertible with its input.
We overcome this shortcoming by gluing together two normalization functions, one sound and one strongly complete, and establishing that the two produce outputs \textbeta\texteta-convertible with each other.
Then, the observation that a strongly complete normalization function that is \textbeta\texteta-convertible with a sound normalization function is necessarily sound and strongly complete will allow us to conclude the correctness for our final normalization function.

\subsection{Sound and Strongly Complete Normalization}
We glue together a sound normalization algorithm and a strongly complete normalization algorithm to establish full correctness for the strongly complete normalization algorithm.

\begin{construction}
    \label{Construction:FinalGluing}
    The gluing \P-category \(\widehat{\mathsf{Subst}}_{\alpha} \downarrow \widehat{\mathsf{Subst}}_{\alpha}\) is a Cartesian-closed \P-category.
    We may therefore induce a Cartesian-pre-closed \P-functor:
    \[ \langle y \downarrow \langle j \rangle j \rangle : \mathsf{Subst}_\alpha \rightarrow \widehat{\mathsf{Subst}}_{\alpha} \downarrow \widehat{\mathsf{Subst}}_{\alpha} \enspace, \]
    where the domain component is the Yoneda embedding, \(y : \mathsf{Subst}_\alpha \rightarrow \widehat{\mathsf{Subst}}_{\alpha}\), and the codomain component is the nerve of \(j\) after \(j\), \(\langle j \rangle \, j : \mathsf{Subst}_\alpha \rightarrow \widehat{\mathsf{Subst}}_{\alpha}\).
    There is a canonical \P-natural transformation, \(y \Rightarrow \langle j \rangle j\), by reflexivity of \textbeta\texteta-conversion.
\end{construction}

\begin{figure}
        \centering
        \begin{tikzcd}
            & {\mathsf{Rnm}}
                \arrow[rd, start anchor=south east, "{i}"{above}, ""{name=A, below}]
                \arrow[ld, start anchor=south west, "{i}"{above}] \\
        	{\mathsf{Subst}_\alpha}
                \arrow[d, "{j}"{left}]
            && {\mathsf{Subst}_\alpha}
                \arrow[d, "{\langle y \downarrow \langle j \rangle j \rangle}"{right}] \\
        	{\mathsf{Subst}_{\beta\eta}}
                \arrow[rr, "{\llbracket - \rrbracket}"{below}]
                \arrow[Rightarrow, from=A, shift right=2, "{\tilde{u}}"{above}, shorten <=3ex, shorten >=3ex]
                \arrow[Leftarrow, from=A, shift left=2, "{\tilde{q}}"{below}, shorten <=3ex, shorten >=3ex]
            && {\widehat{\mathsf{Subst}}_{\alpha} \downarrow \widehat{\mathsf{Subst}}_{\alpha}}
        \end{tikzcd}
        \vspace{2ex}
        \caption{Interpretation of \texttt{Subst}\textsubscript{\textbeta\texteta} into the Gluing \P-Category \(\widehat{\mathsf{Subst}}_{\alpha} \downarrow \widehat{\mathsf{Subst}}_{\alpha}\)}
        \label{Figure:Norm4}
    \end{figure}
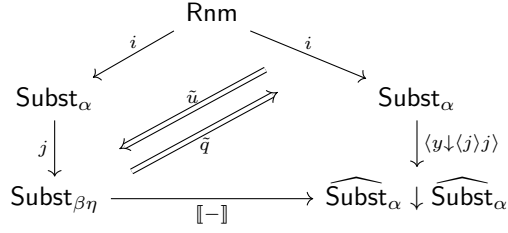

\begin{definition}[\texttt{nf}\textsubscript{4}]
    \label{Definition:FinalNF}
    We define two normalization functions thus:
    \begin{itemize}
        \item \(\mathsf{nf}^\mathrm{D}_4(\sigma : \Gamma \rightarrow_{\mathrm{sub}} \Delta) \triangleq \mathrm{Dom}(\tilde{q}_{\Delta})_{\Gamma}(\mathrm{Dom}(\llbracket \Gamma \vdash_\mathrm{sub} \sigma : \Delta \rrbracket)_\Gamma(\mathrm{Dom}(\tilde{u}_{\Gamma})_{\Gamma}(\mathrm{id}_\Gamma)))\); and
        \item \(\mathsf{nf}^\mathrm{C}_4(\sigma : \Gamma \rightarrow_{\mathrm{sub}} \Delta) \triangleq \mathrm{Cod}(\tilde{q}_{\Delta})_{\Gamma}(\mathrm{Cod}(\llbracket \Gamma \vdash_\mathrm{sub} \sigma : \Delta \rrbracket)_\Gamma(\mathrm{Cod}(\tilde{u}_{\Gamma})_{\Gamma}(\mathrm{id}_\Gamma)))\).
    \end{itemize}
    based on figure~\ref{Figure:Norm4}.
    \end{definition}

\begin{theorem}[Soundness]
    \label{Theorem:Soundness4}
    The normalization function \(\mathsf{nf}^\mathrm{C}_4\) is sound:
    \[ \sigma \sim_{\beta\eta} \mathsf{nf}^\mathrm{C}_4(\sigma) \enspace. \]
\end{theorem}

\begin{theorem}[Strong Completeness]
    \label{Theorem:StrongCompleteness4}
    The normalization function \(\mathsf{nf}^\mathrm{D}_4\) is strongly complete:
    \[ \sigma \sim_{\beta\eta} \sigma' \Rightarrow \mathsf{nf}^\mathrm{D}_4(\sigma) \equiv_\alpha \mathsf{nf}^\mathrm{D}_4(\sigma') \enspace. \]
\end{theorem}

\begin{lemma}
    The normalization functions \(\mathsf{nf}^\mathrm{D}_4\) and \(\mathsf{nf}^\mathrm{C}_4\) agree extensionally:
    \[ j(\mathsf{nf}^\mathrm{D}_4(\sigma)) \sim_{\beta\eta} \mathsf{nf}^\mathrm{C}_4(\sigma) \enspace. \]
\end{lemma}

\begin{theorem}[Correctness]
    \label{Theorem:Correctness4}
    The normalization function \(\mathsf{nf}^\mathrm{D}_4\) is sound and strongly complete:
    \begin{itemize}
        \item \(\sigma \sim_{\beta\eta} j(\mathsf{nf}^\mathrm{D}_4(\sigma))\); and
        \item \(\sigma \sim_{\beta\eta} \sigma' \Rightarrow \mathsf{nf}^\mathrm{D}_4(\sigma) \equiv_\alpha \mathsf{nf}^\mathrm{D}_4(\sigma')\).
    \end{itemize}
\end{theorem}

We have therefore succeeded in categorically constructing a sound and strongly complete normalization function.
That these properties have been induced purely by categorical universal property relies strongly on our novel definitions of Cartesian-pre-closure and the universal property of \texttt{Subst}\textsubscript{\textalpha} relative to \texttt{Subst}\textsubscript{\textbeta\texteta}.

\begin{remark}
    For presentational simplicity, in construction~\ref{Construction:FinalGluing} and definition~\ref{Definition:FinalNF} we used \({\widehat{\mathsf{Subst}}_{\alpha} \downarrow \widehat{\mathsf{Subst}}_{\alpha}}\) as our gluing \P-category, which is simply the arrow category over \(\widehat{\mathsf{Subst}}_{\alpha}\) (and therefore also a presheaf \P-category).
    In fact, it suffices to use any appropriate gluing \P-category of the form \(\widehat{\mathbb{C}} \downarrow F\), where \(\mathbb{C}\) is a \P-category of contexts supporting normalization, and \(F\) is an appropriate \P-functor from any Cartesian-closed \P-category, supporting soundness, into \(\widehat{\mathbb{C}}\).
    Thus, instead, one may use \(\widehat{\mathsf{Rnm}} \downarrow \langle j \circ i \rangle\) and induce the interpretation by \(\langle \langle i \rangle \downarrow j \rangle\).
    This, and related choices, brings our construction closer into connection with the gluing construction of \cite{fiore_2002,fiore_2022}.
\end{remark}

\subsection{Examples}
To demonstrate the normalization algorithm synthesized, we give a few examples of the computation of \(\mathsf{nf}\) on sample terms.

\begin{definition}[\texttt{one}]
    Define \(\mathsf{one}\) as a Church numeral:
    \begin{align*}
        \mathsf{one} \triangleq & \, \mathsf{Abs\,(Var\,IdxZero)}
        \ : \enspace
        \mathsf{Tm\,Nil\,(Arr\,(Arr\,Iota\,Iota)\,(Arr\,Iota\,Iota))} \enspace.
    \end{align*}
\end{definition}

\begin{example}[\texteta-Expansion of \texttt{one}]
    The definition of \(\mathsf{one}\) is not in \texteta-long form.
    The normalization algorithm \texteta-expands it thus:
    \[
        \mathsf{nf}(\mathsf{one})
        \,\equiv\,
        \mathsf{(Abs\,(Abs\,(App} \mathsf{(Var\,(IdxSucc\,IdxZero))}  \mathsf{(Var\,IdxZero))))} \enspace.
    \]
\end{example}

\begin{definition}[\texttt{succ}]
    Define \(\mathsf{succ}\) as the successor Church numeral:
    \begin{align*}
        \mathsf{succ} \triangleq & \, \mathsf{Abs\,(Abs\,(Abs\,(App} \\ & \qquad \,\, \mathsf{(Var\,(IdxSucc\,IdxZero))} \\ & \qquad \,\, \mathsf{(App\,(App} \\ & \qquad \qquad \mathsf{(Var\,(IdxSucc\,(IdxSucc\,IdxZero)))} \\ & \qquad \qquad \mathsf{(Var\,(IdxSucc\,IdxZero)))} \\ & \qquad \quad \, \, \mathsf{(Var\,IdxZero)))))} \enspace.
    \end{align*}
\end{definition}

\begin{example}[Normalization of \texttt{two}]
    The definition of \(\mathsf{two}\) as the successor of \(\mathsf{one}\) is not a \textbeta-normal form.
    It is normalized thus:
    \begin{align*}
        \mathsf{nf}(\mathsf{App\,succ\,one}) \equiv\; & \mathsf{(Abs\,(Abs\,(App} \\ & \quad \mathsf{(Var\,(IdxSucc\,IdxZero))} \\ & \quad \mathsf{(App\,(Var\,(IdxSucc\,IdxZero))\,(Var\,IdxZero)))))} \enspace.
    \end{align*}
\end{example}

\section{Conclusions}
\label{Section:Conclusions}

We conclude by summarising our results and proposing some directions of future work.

\subsection{Summary of Results}
We have taken the pen-and-paper argument of \cite{cubric_normalization_1998} and have formalized it in the computational setting of the \coq{} proof assistant.
The facility of this being achieved was stressed in their paper, and we have demonstrated it.
We have bypassed their non-categorical proof of strong completeness, and instead provided a categorical one by way of a new universal property for unquotiented syntax and categorical gluing.

Our formalization of their work has resulted in the formalization of both \P-category theory and the simply typed \textlambda-calculus.
We have formalized results not known to have been formalized before connecting syntax with both categorical and computational semantics.

Our work fully connects the normalization of simply typed \textlambda-calculus terms with the categorical semantics of the simply typed \textlambda-calculus.
It differs from much other work in that it successfully uses the formalization of abstract category theory for a concrete computational purpose: the normalization of simply typed \textlambda-calculus terms.

\subsection{Further Work}
We suggest a few avenues of further work, not remarked throughout the course of the paper.

We wonder whether the universal property of \texttt{Subst}\textsubscript{\textalpha} relative to \texttt{Subst}\textsubscript{\textbeta\texteta} (theorem~\ref{Theorem:RelativeUniversalProperty}) could be generalized; specifically, we were unable to prove any stronger \P-naturality properties for \(\tilde{q}\) and \(\tilde{u}\).

The use of gluing categories in the final construction connects with the works of \cite{fiore_2002,fiore_2022} and \cite{altenkirch_norm_1995}.
Further analysis should investigate this connection.

Our concept of Cartesian-pre-closure can be generalized to other adjunctions and such lax preservation of right adjoints: Does this situation have broader and more general applicability in category theory?

We use a single base type to generate our simple type theory, other generating data may also be considered, \emph{e.g.}, constants and equations.
The product type former can be readily incorporated; other type formers, such as sums (\cite{altenkirch_norm_coprod_2001,balat_ext_norm_2004}), need be considered.

Finally, investigations should be made about how our technique translates into the settings for more sophisticated type theories such as polymorphic and dependent ones.

\section*{Acknowledgements}
We are grateful to Djordje \v{C}ubri\'c and Peter Dybjer for insightful conversation about the background to their original work.
The research of the second author was partially supported by EPSRC grant EP/V002309/1.

\bibliographystyle{msclike}
\bibliography{refs}

\appendix

\section{Selected \coq{} Code Listings}
\numberwithin{listing}{section}
\includecollection{rocq}

\end{document}